\documentclass[journal=jctcce,manuscript=article]{achemso}

\usepackage{chemformula} 
\usepackage[T1]{fontenc} 

\usepackage{graphicx}
\usepackage{dcolumn}
\usepackage{bm}
\usepackage{hyperref}
\usepackage{amssymb}
\usepackage[caption=false]{subfig}
\usepackage{booktabs}
\usepackage{xcolor}

\newcommand{\myciteauthor}[1]{{\protect\NoHyper\citeauthor{#1}\endNoHyper}}
\newcommand{\onlinecite}[1]{\hspace{-1 ex} \nocite{#1}\citenum{#1}}
\DeclareGraphicsExtensions{.pdf,.png,.jpg}

\author{Stefan Riemelmoser}
\affiliation{University of Vienna, Faculty of Physics and Center for Computational Materials Science,  Kolingasse 14-16,
A-1090 Vienna, Austria}
\alsoaffiliation{University of Vienna, Vienna Doctoral School in Physics,  Boltzmanngasse 5,
A-1090 Vienna, Austria
}
\email{stefan.riemelmoser@univie.ac.at}
\author{Carla Verdi}
\affiliation{University of Vienna, Faculty of Physics and Center for Computational Materials Science,  Kolingasse 14-16,
A-1090 Vienna, Austria}
\alsoaffiliation{%
School of Physics, The University of Sydney, Sydney, NSW 2006, Australia 
}%
\author{Merzuk Kaltak}
\affiliation{%
VASP Software GmbH, Sensengasse 8/12,
A-1090 Vienna, Austria 
}
\author{Georg Kresse}
\affiliation{University of Vienna, Faculty of Physics and Center for Computational Materials Science,  Kolingasse 14-16,
A-1090 Vienna, Austria}
\alsoaffiliation{%
VASP Software GmbH, Sensengasse 8/12,
A-1090 Vienna, Austria 
}
\title[Machine learning density functionals]
  {Machine learning density functionals from the random-phase approximation}

\begin{document}

\begin{tocentry}

\includegraphics {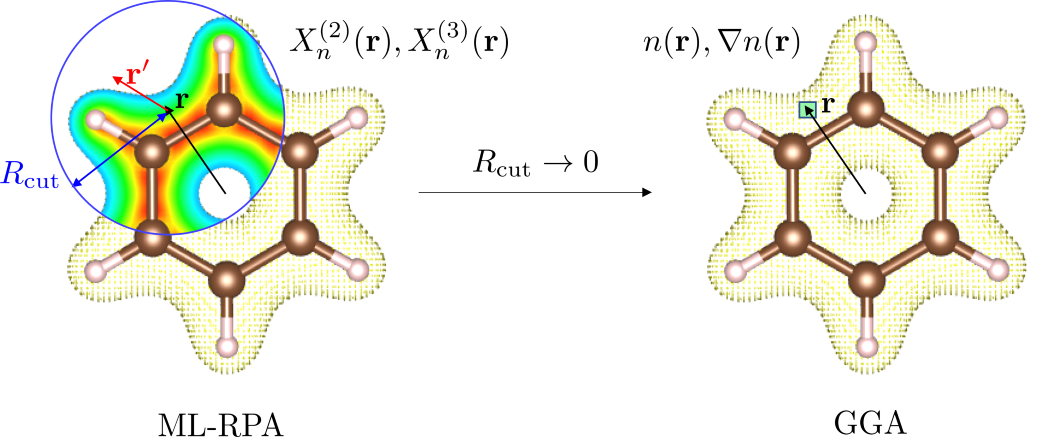}
%

\end{tocentry}

\begin{abstract}
Kohn-Sham density functional theory (DFT) is the standard method for first-principles calculations in computational chemistry and materials science. More accurate theories such as the random-phase approximation (RPA) are limited in application due to their large computational cost. Here, we construct a DFT substitute functional for the RPA using supervised and unsupervised machine learning (ML) techniques.  
Our ML-RPA model can be interpreted as a non-local extension to the standard gradient approximation. We train an ML-RPA functional for diamond surfaces and liquid water and show that ML-RPA can outperform the standard gradient functionals in terms of accuracy. Our work demonstrates how ML-RPA can extend the applicability of the RPA to larger system sizes, time scales and chemical spaces. 
\end{abstract}

\section{Introduction}

For over half a century, generations of researchers have been looking to find ever better approximations for the elusive exchange-correlation (xc) functional of Kohn-Sham density functional theory (DFT).\cite{Burke2012,Medvedev2017} The Hohenberg-Kohn theorems guarantee this exchange-correlation functional to be a universal functional of the electronic density $n$, in other words, there exists a map $n \mapsto E_{\rm xc}[n]$.\cite{Hohenberg1964} However, the exact functional is of complex non-local form and in general unknown.  Common approximations for the exchange-correlation energy are built on the principle of near-sightedness:\cite{Kohn1999} in the local density approximation (LDA),\cite{Kohn1965} the exchange-correlation energy density is approximated to be locally that of a homogeneous electron gas with the same density. 

To improve the accuracy of the approximation, one can add more non-local information by climbing up Jacob's ladder of DFT.\cite{Perdew2001} Next to the local density, one includes its gradient $|\nabla n|$ as ingredient to the exchange-correlation functional (generalized gradient approximation or GGA).  
Going beyond the GGA, more non-local information is added in the form of the kinetic energy density (meta-GGA), and fractions of exact exchange (hybrids). However, strictly speaking, most meta-GGA and hybrid functional approaches deviate from ``pure'' Kohn-Sham DFT, as the ingredients go beyond the electronic density alone.\cite{Seidl1996}
A different route beyond GGA are non-local van der Waals (vdW) functionals,\cite{Dion2004,*Dion2005,Berland2015} which account for pair-wise dispersion interactions between densities $n(\textbf{r})$ and $n(\textbf{r}')$. As the ingredients are only the electronic density and its gradient at points $\textbf{r}$ and $\textbf{r}'$, the non-local vdW method stays within pure KS-DFT.

Given ingredients $\textbf{X}$ for the exchange-correlation functional, one still has to find the actual functional form, that is the map $\textbf{X} \mapsto E_{\rm xc}[\textbf{X}]$. In the analytical approach developed by Perdew and others,\cite{Perdew1996,*Perdew1997,Perdew1996a,Sun2015} those maps are found by satisfying exact constraints. The empirical approach pursued by Becke and others\cite{Becke1988,Lee1988,Becke1993,Zhao2007} optimizes a small number of adjustable parameters to experimental data or higher level theory. Broadly speaking, the analytical functionals are more universally applicable, whereas the empirical approach can achieve higher accuracy for systems similar to the ones represented in their respective training sets.\cite{Medvedev2017} This comes at the cost of worse performance for systems not represented by these training sets. For example, the widely used B3LYP functional,\cite{Stephens1994} whose parameters have been optimized for small main group molecules, performs well for main group chemistry but struggles when applied to extended systems\cite{Paier2007} and transition metal chemistry.\cite{Cramer2009} 

Recently, machine learning (ML) techniques have been taking the empirical approach to its extreme.\cite{Nagai2020,Bogojeski2020,Dick2021,Kirkpatrick2021} Not limited by human parametrizations, machine learning approaches can optimize the maps $\textbf{X}\mapsto E_{\rm xc}[\textbf{X}]$ using complicated non-linear functional forms. The first attempt of creating an ML density functional goes back to \myciteauthor{Tozer1996},\cite{Tozer1996} an effort which culminated in the development of the HTCH functional (Hamprecht-Tozer-Cohen-Handy, a popular GGA functional).\cite{Hamprecht1998}
The full potential of ML approaches has been demonstrated in pioneering work by Burke and coworkers, who showed that orbital-free density functionals can be learned from the full non-local density.\cite{Snyder2012,Li2015} Their approach was later applied to standard Kohn-Sham DFT, enabling molecular dynamics simulations of single molecules with chemical accuracy.\cite{Brockherde2017,Bogojeski2020,Kalita2021} Though this is very impressive, these ML-DFT functionals are tailor-made for this specific purpose and have to be retrained for every new molecule. A different approach was taken by \myciteauthor{Nagai2020},\cite{Nagai2020} who complemented meta-GGA ingredients by a non-local density descriptor, achieving remarkable accuracy for a large molecular test set with training data from only three molecules. For a broader discussion of different ML-DFT approaches, we refer also to the review of Schmidt \textit{et al.}\cite{Schmidt2019a}  
 
In the current work, we propose an approach to construct machine learned density functionals from the random-phase approximation (RPA, a high-level functional from the top of Jacob's ladder\cite{Perdew2001}). We adapt the power spectrum representation of atomic environments used for machine-learned force fields\cite{Bartok2013,*Bartok2013a,*Bartok2017,Jinnouchi2019a} (MLFF) to construct ingredients for ML-DFT. We show that these ingredients can be considered to be a non-local extension of GGA. 
In MLFF, data efficiency can be improved by training not only on energies alone but rather also on atomic forces.\cite{Chmiela2017} Analogously, derivative information in DFT can be supplied via the exchange-correlation potentials,
\begin{equation}\label{eq:xc-potential}
v_{\rm xc}(\textbf{r})=\delta E_{\rm xc} / \delta n(\textbf{r}) .
\end{equation}
However, obtaining accurate exchange-correlation potentials from beyond GGA functionals is generally difficult, and aside from the early work of Tozer \textit{et al.}, this approach has only been applied to simple model systems.\cite{Nagai2018,Zhou2019,Schmidt2019} Here, we supply such derivative information by using our recent implementation of the optimized effective potential method to obtain exchange-correlation potentials from the RPA.\cite{Riemelmoser2021}  We demonstrate our method by fitting ML-RPA to diamond and liquid water and show that it enables larger scale RPA calculations. ML-RPA achieves its speed-up via bypassing the optimized effective potential equation, substituting the complicated RPA exchange-correlation functional with pure KS-DFT. Further, our efficient plane-wave implementation brings the system size scaling of ML-RPA down to that of standard DFT. Finally, our approach enables self-consistent calculations, force and stress predictions, and even molecular dynamics simulations for molecules, solids and their surfaces. 
The rest of this paper is organized as follows. Sec. \ref{sec:Riemelmoser2023_2} introduces the ML-RPA formalism. In Sec. \ref{sec:Riemelmoser2023_3}, we briefly discuss the optimized effective potential method and comment on the issue of electronic self-consistency. Results are presented in Sec. \ref{sec:Riemelmoser2023_4} and discussed in Sec. \ref{sec:Riemelmoser2023_5}. Conclusion are drawn in Sec. \ref{sec:Riemelmoser2023_6}.

\section{ML-RPA formalism}\label{sec:Riemelmoser2023_2}

\subsection{Representation of the electronic density}\label{sec:Representation of the electronic density }

We adapt the power spectrum representation of atomic environments\cite{Bartok2013,*Bartok2013a,*Bartok2017} to electronic densities as follows. The electronic density around each real-space grid point $\textbf{r}$ is expanded into radial basis functions $\phi_{nl}$ (described in Supplementary Sec. S1) times real spherical harmonics $Y_l^m$
\begin{equation}
\label{eq:cnlm}
\begin{aligned}
n(\textbf{r}+\textbf{r}') f_{\rm cut}(r') = \sum_{nlm} c_{nlm}(\textbf{r}) \phi_{nl}(r')  Y_l^m(\widehat{\textbf{r}'}) .
\end{aligned}
\end{equation}
The cutoff function $f_{\rm cut}$ puts emphasis on nearby densities ($r' \leq R_{\rm cut}$), following Kohn's principle of nearsightedness.\cite{Behler2007,Kohn1999} Here, we use a cutoff radius of $R_{\rm cut}=1.5 \text{ \AA}$.\\
The expansion coefficients $c_{n00}$ are the equivalents of rotationally invariant two-body descriptors used in MLFF, thus we write $X_n^{(2)}=c_{n00}$.
In the limit of small cutoffs $R_{\rm cut}$, the $X_n^{(2)}$ reduce to the local density
\begin{equation}\label{eq:2-body limit}
\begin{aligned}
X_n^{(2)}(\textbf{r}) \propto n(\textbf{r}) + \mathcal{O}(R_{\rm cut}^2) \hspace{14pt} \text{for } R_{\rm cut} \to 0 ,
\end{aligned}
\end{equation}
as shown in Supplementary Sec. S1.
It is interesting that the well-known weighted density approximation\cite{Gunnarsson1979} can be viewed as the special case where only a single two-body descriptor is taken, compare also the non-local density descriptor introduced by Nagai \textit{et al.}\cite{Nagai2020} 

Further, angular information is accounted for by forming rotationally invariant combinations of the $l=1$ expansion coefficients to construct additional density descriptors $X_n^{(3)}$, similar to the three-body descriptors in MLFF,
\begin{equation}\label{eq:3-body descriptors}
X_n^{(3)} = \frac{\sigma^{(3)}}{R_{\rm cut}}\sqrt{c_{n1x}^2+c_{n1y}^2+c_{n1z}^2}.
\end{equation}
Here, $\sigma^{(3)}$ is an ML hyperparameter that weighs the $X^{(3)}_n$ relative to the $X_n^{(2)}$.\cite{Jinnouchi2020}
For small cutoffs, the $X_n^{(3)}$ reduce to the local gradient
\begin{equation}\label{eq:3-body limit}
X_n^{(3)}(\textbf{r}) \propto |\bm{\nabla} n (\textbf{r}) |+ \mathcal{O}(R_{\rm cut}^2) \hspace{14pt} \text{for } R_{\rm cut} \to 0 ,
\end{equation}
as demonstrated in Supplementary Sec. S1. In summary, an exchange-correlation functional with $X_n^{(2)}$ and $X_n^{(3)}$ as its ingredients can be considered as a \textit{non-local extension of the generalized gradient approximation}. Here, we use 4 radial basis functions, thus in total we have $4+4=8$ density descriptors.
A representation based on non-local convolutions of the electronic density has been previously suggested by \myciteauthor{Lei2019},\cite{Lei2019} who also showed how the representation can be systematically completed by adding higher body-order descriptors. Likewise, representations of atomic environments for MLFF can be completed via moment tensor potentials\cite{Shapeev2016} or the atomic cluster expansion.\cite{Drautz2019,*Drautz2019a} Next to density convolutions proposed by \myciteauthor{Lei2019},\cite{Lei2019} descriptors similar to ours have been used also in several other recent works, see Refs. \onlinecite{Grisafi2018,Margraf2021,Dick2020,Bystrom2022}. A key distinction is that some of these works use density fitting to construct descriptors for the electronic density. As pointed out by \myciteauthor{Chen2020},\cite{Chen2020} the explicit dependence on chemical species makes those ML-DFT functionals less universal.

\subsection{Machine learning scheme}
The exchange-correlation energy for any GGA functional can be written as
\begin{equation}
E_{\rm xc}^{\rm GGA} = \int \text{d} \textbf{r}\; n( \textbf{r}) \varepsilon_{\rm x,HEG}[n(\textbf{r})] F_{\rm xc}[n(\textbf{r}),|\bm{\nabla} n (\textbf{r})|],
\end{equation}
where $\varepsilon_{\rm x, HEG}$ is the exchange energy density for the homogeneous electron gas of uniform density $n$, and $F_{\rm xc}$ is the enhancement factor. We use the same form as an ansatz for our ML-RPA model, 
\begin{equation}\label{eq:ML-RPA ansatz}
E_{\rm xc}^{\rm ML-RPA} = \int \text{d} \textbf{r}\; n( \textbf{r}) \varepsilon_{\rm x, HEG}[n(\textbf{r})] F_{\rm xc}^{\rm ML-RPA}[\textbf{X} (\textbf{r})] ,
\end{equation}
where $\textbf{X}$ is a supervector containing the two- and three-body descriptors. The map $ \textbf{X}(\textbf{r}) \mapsto F_{\rm xc}^{\rm ML-RPA}(\textbf{r})$ is found by kernel regression using a Gaussian kernel, 
\begin{equation}\label{eq:Gaussian kernel}
F_{\rm xc}^{\rm ML-RPA}[\textbf{X}(\textbf{r})] = \sum_{i_B} w_{i_B}\exp\left\{-\frac{[\textbf{X}(\textbf{r})-\textbf{X}^{i_B}]^2}{2\sigma^2}\right\} .  
\end{equation}
Here, the kernel width $\sigma$ is an ML hyperparameter, and the $\textbf{X}^{i_B}$ are representative kernel control points chosen from the training data. The corresponding weights $w_{i_B}$ are found by solving a linear regression problem. That is, the ML-RPA exchange-correlation energies and ML-RPA exchange-correlation potentials have to agree with exact RPA reference data in a least square sense. The ML-RPA exchange-correlation potentials are obtained by inserting the ansatz \eqref{eq:ML-RPA ansatz} into Eq. \eqref{eq:xc-potential} and applying the chain rule, see Supplementary Sec. S2. 

\begin{figure}[!!tb]
\centering
\includegraphics [width=0.75\linewidth,keepaspectratio=true] {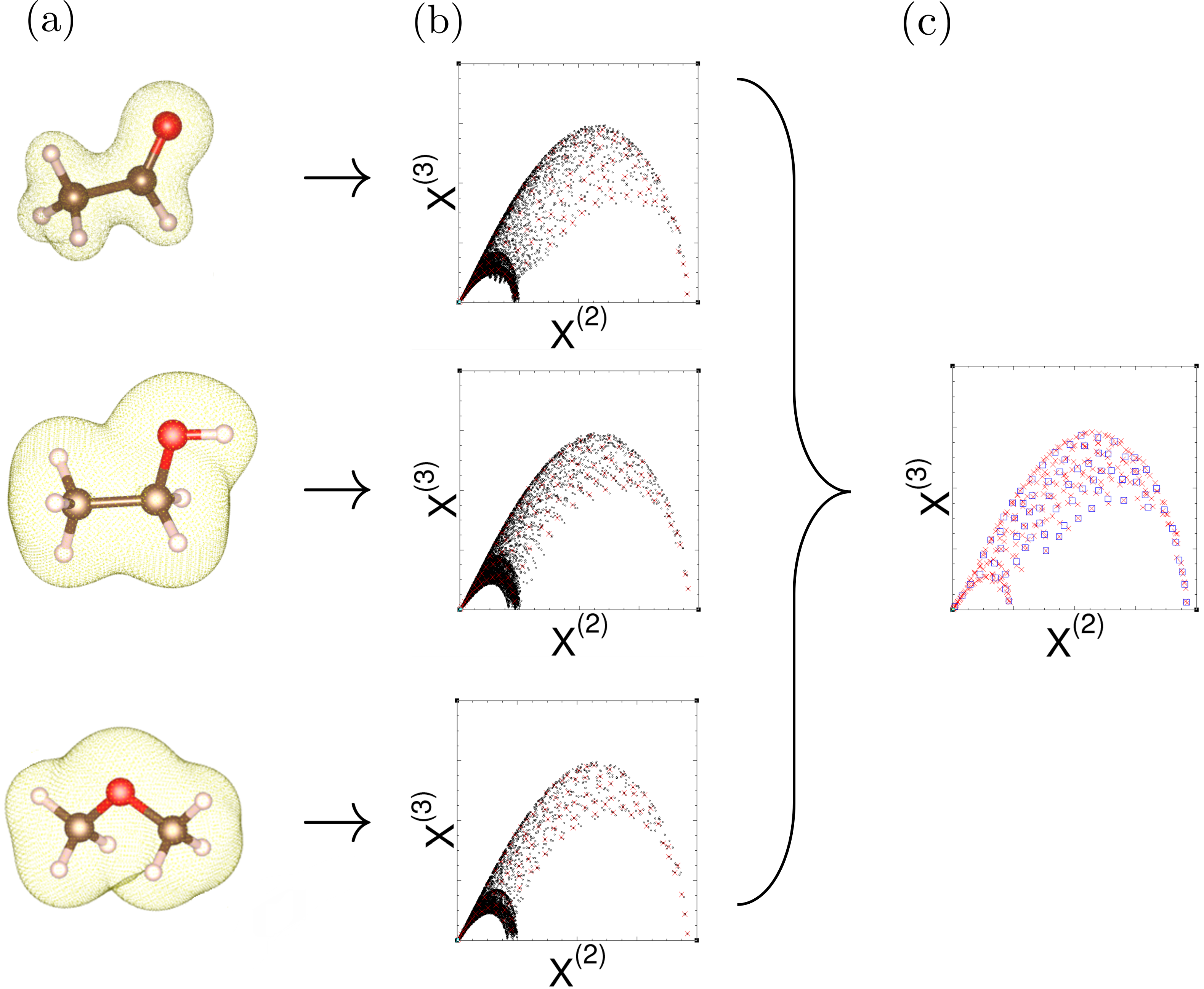}
\caption{Data sparsification scheme for a training set consisting of three molecules, depicted in column (a).  First, using the electronic density as input, we calculate the two-body descriptors $X_n^{(2)}(\textbf{r})$ and three-body descriptors $X_n^{(3)}$(\textbf{r}) at each real-space grid point [black dots in column (b)]. A single radial basis function is used here for visualization purposes. Next, a metric is introduced via the Gaussian kernel and representative density environments (red crosses) are chosen via k-means clustering for each training structure separately. The chosen points are concatenated and a second k-means layer selects the kernel control points $\textbf{X}^{i_B}$ used in Eq. \eqref{eq:Gaussian kernel} [blue squares in column (c)]. For more details, see Supplementary Sec. S3.}
\label{fig:sparsification}
\end{figure}
The exchange-correlation potentials provide derivative information for the ML fit as atomic forces do in MLFF, thus improving data efficiency. That is, instead of a single data point per structure for $E_{\rm xc}^{\rm RPA}$, we have additional data for $v_{\rm xc}^{\rm RPA}(\textbf{r})$ on typically $\mathcal{O}(10^5)-\mathcal{O}(10^6)$ grid points $\textbf{r}$. As this large amount of data cannot be handled by kernel methods, we have to employ data sparsification tools. Without loss of accuracy, the training data can be compressed dramatically by combining k-means clustering\cite{Arthur2006} with the metric induced by the Gaussian kernels (see Supplementary Sec. S3). This is illustrated in Fig. \ref{fig:sparsification} using a single radial basis function, such that the descriptors can be easily visualized in 2D. Once training is complete, the computational cost of evaluating ML-RPA depends linearly on the number of kernel control points used. Further, efficient evaluation of the ML-RPA functional is achieved by extensive use of fast Fourier transforms.  We find that ML-RPA is slower than standard GGA functionals by some prefactor for typical applications, but ML-RPA scales as $\mathcal{O}(N\log N)$ with system size $N$ like GGA does.\cite{White1994} In comparison, non-local vdW functionals can have the same system size scaling and similar computational cost as ML-RPA,\cite{RomanPerez2009}
whereas exact RPA is several orders of magnitude slower and has at least $\mathcal{O}(N^3)$ scaling.\cite{Rojas1995,Kaltak2014}  

\section{Methods}\label{sec:Riemelmoser2023_3}

\subsection{Random-phase approximation}
As the RPA includes information from unoccupied orbitals as ingredients, it sits on the fifth (highest) rung of Jacob's ladder. In the last two decades, the RPA has been successfully employed for a multitude of problems, see Refs. \onlinecite{Ren2012} and \onlinecite{Eshuis2012} for reviews. In particular, the RPA is considered a ``gold standard'' for first-principles surface studies\cite{Schimka2010,Patra2017,Brandenburg2019} due to its seamless inclusion of vdW interactions next to the good description of covalent and metallic bonds. Combining exact exchange and a good description of dispersion interactions makes the RPA also suitable for water and ice.\cite{Macher2014,Yao2021} 
The usual expression for the RPA exchange-correlation energy reads\cite{Langreth1977} 
\begin{equation}
E_{\rm xc}^{\rm RPA} = \text{Tr}[\ln(1-\chi_0 V)],
\label{eq:RPA}
\end{equation}
where $\chi_0$ is the Kohn-Sham response function, $V$ is the Coulomb kernel, and we have used a symbolic notation for sake of brevity. 
As in standard Kohn-Sham DFT, one can obtain the exchange-correlation potential corresponding to the RPA exchange-correlation energy by taking the functional derivative with respect to the electronic density,
\begin{equation}
v_{\rm xc}^{\rm RPA}(\textbf{r})=\delta E_{\rm xc}^{\rm RPA} / \delta n(\textbf{r}) .
\end{equation}
Plugging in Eq. \eqref{eq:RPA} and using the chain rule, one can derive the so-called optimized effective potential (OEP) equation,\cite{Sham1983,Sham1985,Goerling1994,Niquet2003} symbolically
\begin{equation}
\begin{aligned}
\chi_0 v_{\rm xc} = G_0 \Sigma_{\rm xc}^{G_0W_0} G_0 ,
\end{aligned}
\end{equation}
where $G_0$ is the non-interacting (Kohn-Sham) Greens-function and $\Sigma_{\rm xc}^{G_0W_0}$ is the self-energy in the $G_0W_0$ approximation. For more details on the implementations of the RPA exchange-correlation energies and the RPA-OEP method, we refer to Refs. \onlinecite{Kaltak2014} and \onlinecite{Riemelmoser2021}, respectively. 

\subsection{Electronic self-consistency}

\begin{figure}[!htb]
\centering
\includegraphics [width=0.75\linewidth,keepaspectratio=true] {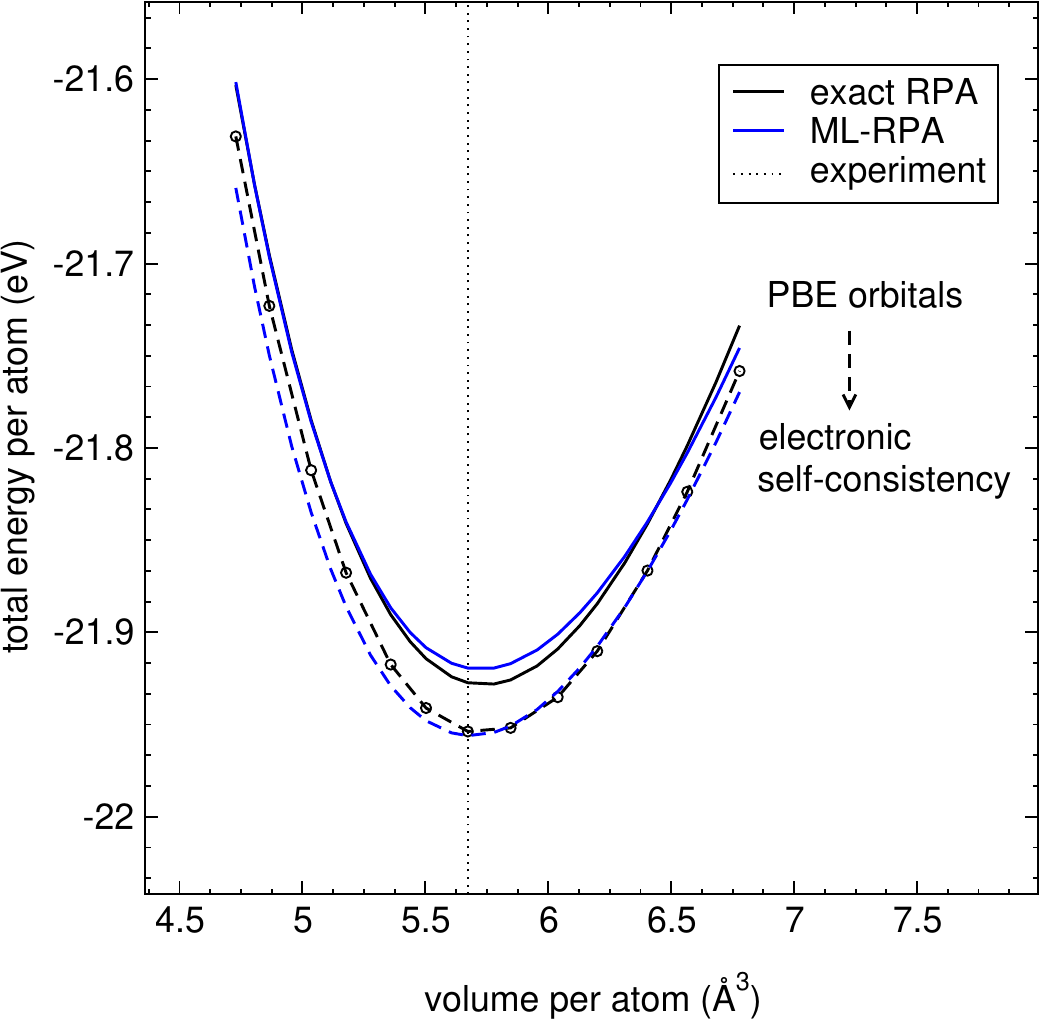}
\caption{Energy-volume curves of bulk diamond obtained using exact RPA and ML-RPA. Full lines indicate calculations on-top of PBE orbitals, and dashed lines indicate self-consistent calculations. Self-consistent RPA reference energies are obtained using the RPA-OEP method. The dotted vertical line indicates the experimental equilibrium volume.\cite{Warren1967}}
\label{fig:diamond_volume}
\end{figure}

Even tough the RPA-OEP method allows in principle to perform RPA calculations self-consistently,\cite{Riemelmoser2021} this procedure is seldom carried out due to the large computational overhead. A more common approach is to evaluate the RPA using orbitals from a semi-local DFT base functional (``RPA on-top of DFT'', RPA@DFT).\cite{Harl2010} Here, we calculate all RPA reference data using PBE as base functional (Perdew-Burke-Ernzerhof, a popular GGA functional\cite{Perdew1996,*Perdew1997}). That is, \textit{RPA@PBE is the ground truth for ML-RPA}, and we denote this ground truth in the following as ``exact RPA''. Unless stated otherwise, all RPA calculations and ML-RPA calculations are thus performed non-self-consistently on-top of PBE orbitals. A ML substitute functional that reproduces only RPA@PBE would already be very useful, we reiterate that RPA@PBE is standard practice. 

For the calculation of atomic forces, however, it is convenient to perform self-consistent calculations to avoid tedious non-Hellman-Feynman terms. Wherever exact RPA reference forces are required, for instance for phonon calculations, we calculate such non-Hellman-Feynman terms explicitly.\cite{Ramberger2017} In contrast, ML-RPA can be simply run self-consistently like any semi-local density functional since the exchange-correlation potential is readily available. As a validation, we used the computationally demanding RPA-OEP method to calculate the equilibrium volume of bulk diamond self-consistently. Fig. \ref{fig:diamond_volume} shows that ML-RPA reproduces the RPA@PBE ground truth well (solid lines). Further, ML-RPA correctly predicts a small downward shift due to electronic self-consistency (dashed lines).

We would like to emphasize that this result is not obvious, as the training set covers only PBE densities. To this point, \myciteauthor{Snyder2012}\cite{Snyder2012} have reported early on that electronic self-consistency can grossly deteriorate the performance of ML-DFT functionals, as self-consistency leads the functional away from its training manifold. Likewise, we observed that earlier versions of ML-RPA became inaccurate when applied self-consistently. However, we find that the current ML-RPA is very stable and reliably converges to a tight energy threshold of $10^{-8}$ eV. Key hyperparameters in this regard are the cutoff radius ($R_{\rm cut} = 1.5 \text{ \AA}$) and the Tikhonov regularization parameter ($t_{\rm SVD} = 1.0\times 10^{-9}$, see Supplementary Sec. S2). Even atomic densities can be used as starting points, tough preconverging with PBE typically speeds up the ML-RPA self-consistency cycle.  Lastly, the ML-RPA stress tensor has also been implemented via finite differences, which is useful for example for volume relaxations and the training of machine learned force fields (see Supplementary Sec. S5).

\subsection{Computational details}

We use the  PAW code \texttt{VASP} (Vienna \textit{Ab Initio} Simulation Package),\cite{Kresse1996} adopting the C\_GW, H\_GW and O\_GW pseudo-potentials. All DFT and RPA calculations are performed spin-non-polarized. An energy cutoff of 600 eV is used for the plane-wave orbital basis set (\texttt{ENCUT} in VASP). For RPA calculations, a reduced cutoff of 400 eV is used to expand the response function $\chi_0$ (\texttt{ENCUTGW} in \texttt{VASP}), using a cosine window to smoothen the cutoff of the Coulomb kernel.\cite{Harl2010,Riemelmoser2020} Basis set incompleteness errors are discussed in Supplementary Sec. S4. The one-center PAW contributions to the RPA exchange-correlation energy are treated on the level of Hartree-Fock, consistently for RPA and ML-RPA calculations. The good agreement of \textit{total energies} in Fig. \ref{fig:diamond_volume}, rather than relative energies only, demonstrates the consistency of the ML-RPA implementation. Similar agreement is observed for all materials, with fit errors around 1 meV/electron.

\section{Results}\label{sec:Riemelmoser2023_4}

\subsection{Bulk diamond}

\begin{figure}[!htb]
\centering
\subfloat{\includegraphics [angle=-90,width=0.55\linewidth,keepaspectratio=true] {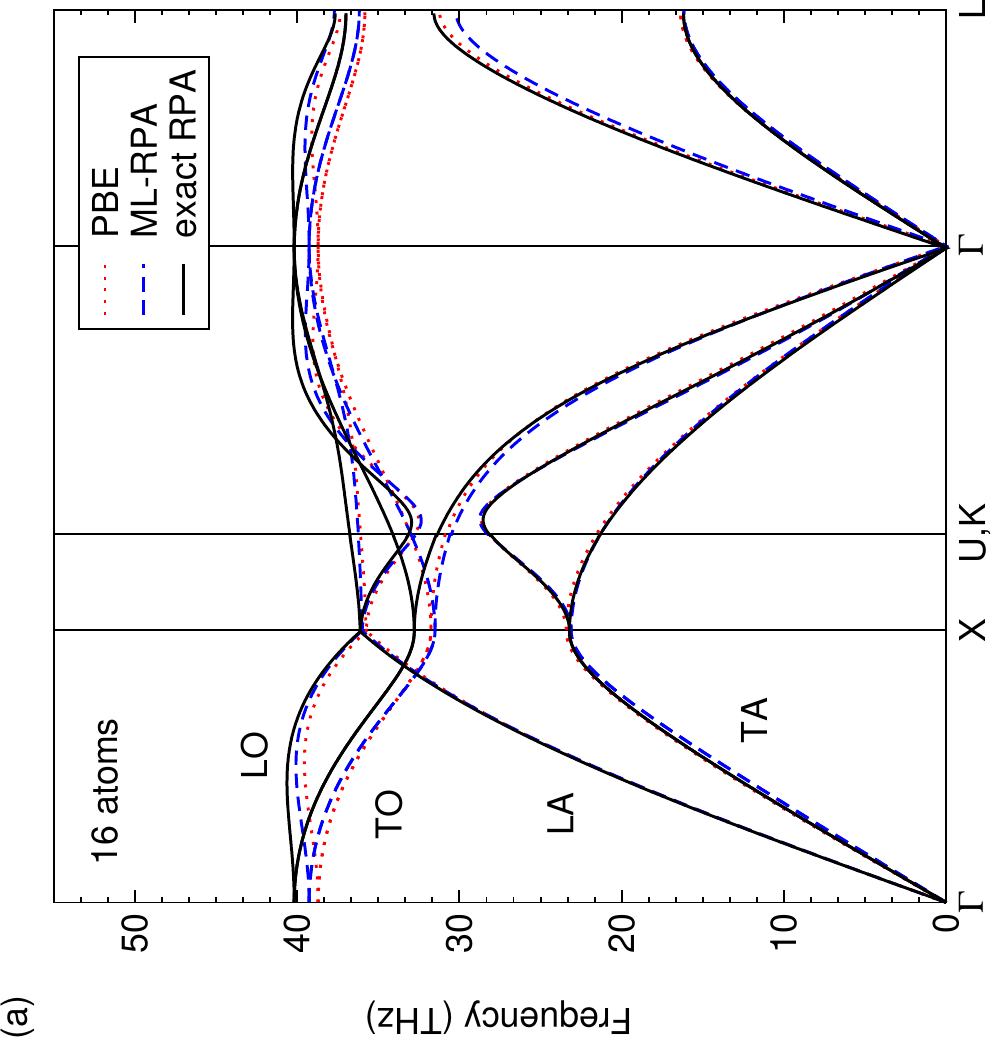}\label{fig:diamond_phonons_16}}
\par\medskip
\subfloat{\includegraphics [angle=-90,width=0.55\linewidth,keepaspectratio=true] {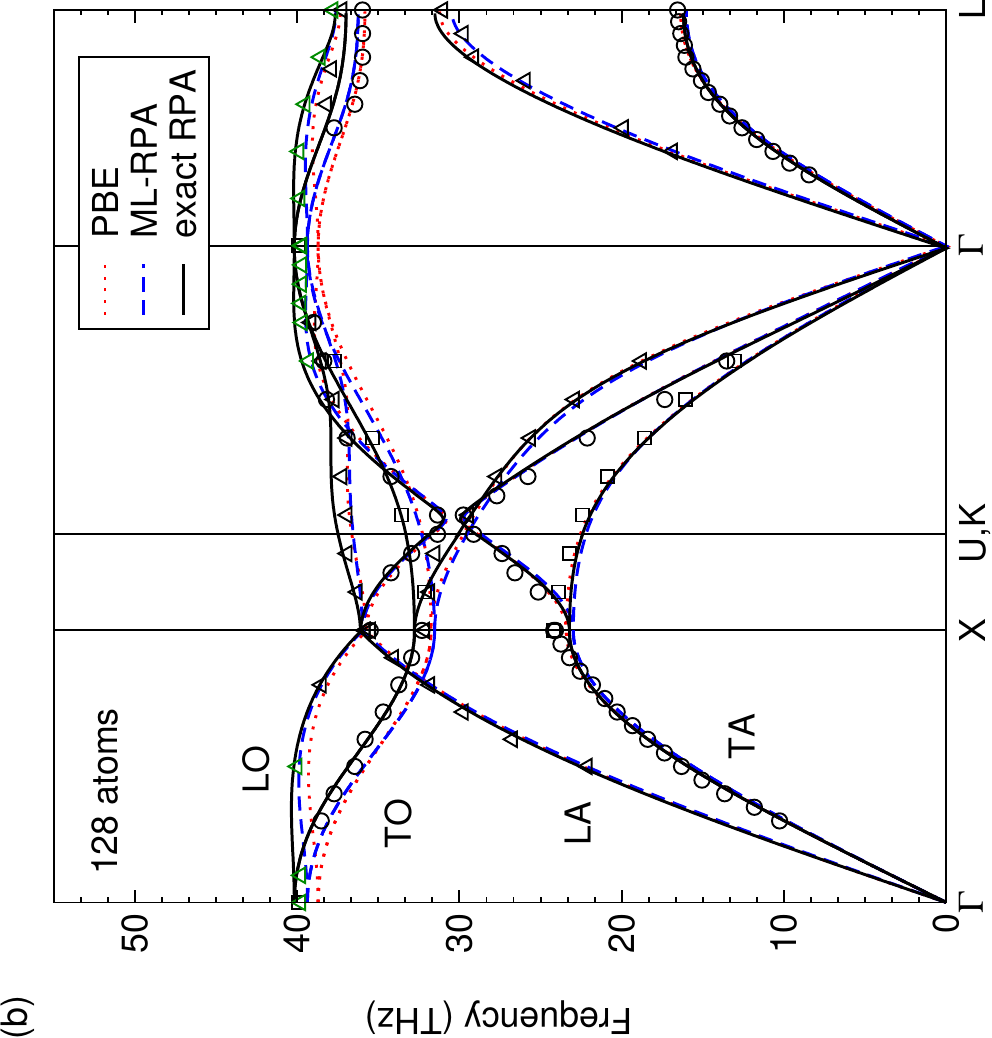}\label{fig:diamond_phonons_128}}
\caption{Phonon dispersion of diamond, calculated at the respective equilibrium lattice constants, using supercells containing (a) 16 atoms, (b) 128 atoms. Phonons obtained using exact RPA are represented as solid black lines, blue dashed lines represent ML-RPA. PBE calculations are also shown for comparison (red dotted lines). ML-RPA has training data only for the smaller supercell size [16 atoms, panel (a)]. Black and green symbols indicate experimental data from Refs. \protect\onlinecite{Warren1967} and \protect\onlinecite{Kulda2002}, respectively.}
\label{fig:diamond phonons}
\end{figure}

We apply our ML scheme to create an RPA substitute functional for diamond and liquid water.
Following a common MLFF practice,\cite{Behler2007} bulk diamond training data are iteratively added from molecular dynamics (MD) simulations, using prior versions of the ML-RPA functional to create the MD trajectories. We pick 10 MD snapshots from the 8 atom supercell and 5 snapshots from the 16 atom supercell. The full ML-RPA training set is further detailed in Supplementary Sec. S4. The equilibrium lattice constant obtained by ML-RPA (3.582 \AA) agrees well with the exact RPA value (3.581 \AA). \\ 
Next, we calculate phonon dispersions using the finite displacement method.\cite{Kresse1995,Engel2020} Converged results are obtained in the large supercell limit, which is hard to achieve with exact RPA due to its unfavorable scaling behavior. Fig. \ref{fig:diamond phonons} compares the ML-RPA phonon dispersions for different supercell sizes to exact RPA results. For contrast, phonon dispersions obtained using PBE are shown as well. For the 16 atom supercell [panel (a)], where training data are available, the ML-RPA phonon dispersion is generally in good agreement with exact RPA, though high-frequency modes are slightly underestimated. Exact RPA calculations for the larger 128 atom supercells validates the extrapolation ability of ML-RPA, as no training data are included for this supercell size. A prominent finite size effect is the closing of the gap near the K-point with respect to the smaller 16 atom cell. Further, the overbending of the LO modes reduces with increasing supercell size, which is most notable along $\Delta$ ($\Gamma$-X). These characteristic features are well reproduced with ML-RPA. The phonon dispersions obtained using exact RPA are overall in good agreement with experimental data,\cite{Warren1967,Kulda2002} whereas ML-RPA and PBE slightly underestimate the high-frequency modes.

\subsection{Diamond surfaces}

\begin{figure}[!htb]
\centering
\includegraphics [width=0.75\linewidth,keepaspectratio=true] {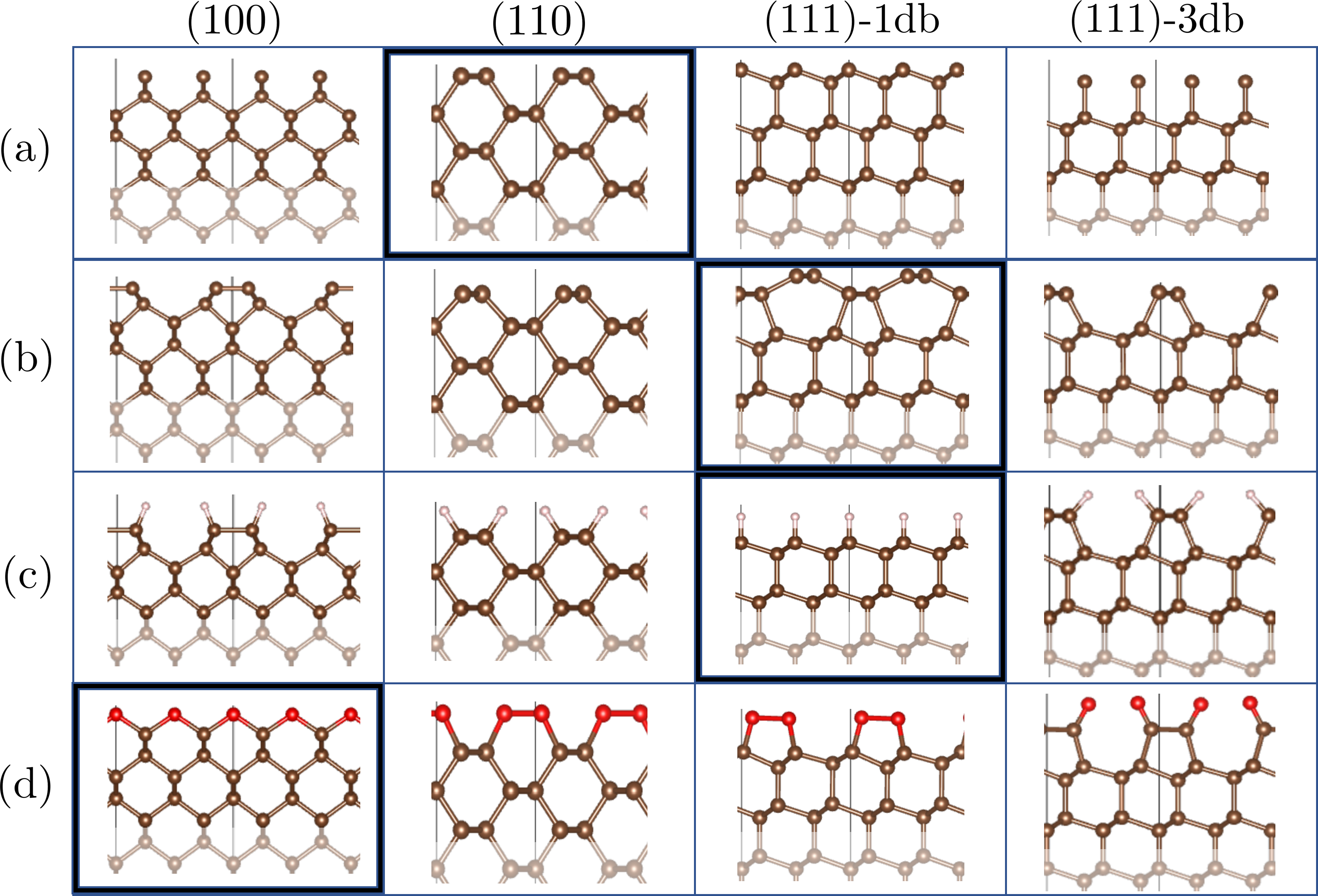}
\caption{Side view of crystallographic diamond surfaces.\cite{Momma2011} 
The $2\times1$ simulation cells are indicated by black vertical lines, and bulk regions are indicated by faded atoms. From top to bottom: (a) clean surfaces, (b) reconstructed surfaces, (c) hydrogenated surfaces (1 ML), (d) oxygenated surfaces (1 ML). Bold edges indicate the most stable orientation for a given surface termination as predicted by exact RPA. Whereas the (111)-1db surface dereconstructs upon the chemisorption of hydrogen and oxygen,\cite{Loh2002} the (100) surface dereconstructs only upon the chemisorption of oxygen.\cite{Sque2006} The (111)-3db surface retains the $2\times1$ geometry throughout,\cite{Kern1996a}  and the (110) surface shows no $2\times1$ reconstruction at all.\cite{Kern1997} }
\label{fig:surfaces}
\end{figure}

The advent of chemical vapor deposition (CVD) has encouraged detailed first-principles simulations of diamond surfaces. 
Specifically, the characterization of ideal crystallographic surfaces has proven useful for the theoretical understanding of CVD grown diamond.\cite{Ristein2006}
When a diamond surface is cut, the outer surface layers can rearrange to partially eliminate the exposed dangling bonds, see Fig. \ref{fig:surfaces}. These reconstructions significantly change the properties of the surfaces. Further, important material properties such as electron affinity can be modified via chemisorption processes in a controlled fashion, the most important surface adsorbates being H and O. \\
The (100) surface is the most relevant crystallographic surface and has been thoroughly studied.\cite{Furthmueller1996,Sque2006} While the (111) surface has also received a lot of attention,\cite{Kern1996,Loh2002} most studies neglect that there are two possible ways to cut this surface.\cite{Zheng1992,Kern1996a} Namely, via the so-called glide and shuffle planes, one can expose (111) diamond surfaces with one dangling bond (1db) or three dangling bonds (3db) per surface atom, respectively. Even though the clean 1db surface is clearly more stable, 3db surfaces naturally occur in growth and etching processes. Finally, the (110) surface, which is difficult to prepare experimentally, has only recently been fully characterized.\cite{Chaudhuri2022} \\
In the following, we calculate diamond surface energies of formation via
\begin{equation}
E_{\rm surf} = (E - E_{\rm dia}/N_{\rm dia} - E_{\text{H}_2}/N_{\text{H}}-E_\text{O}/N_\text{O})/N_{\rm surf},
\label{eq:formation-surface}
\end{equation}
where $E$ is the total energy of the surface, $E_{\rm dia}$ and $E_{\text{H}_2}$ are the energies of bulk diamond and the $\text{H}_2$ molecule, and $E_\text{O}$ is calculated using the water monomer as reference 
\begin{equation}
E_\text{O} = E_{\text{H}_2\text{O}}-E_{\text{H}_2} .
\end{equation}
Vibrational contributions to the formation energies due to zero-point motion are neglected.
Diamond surface calculations are performed in $2\times1$ supercells using symmetric slabs. We use 18 surface layers for the (111)-1db surface and 16 layers for the rest. This assures convergence of the surface formation energies to better than 10 meV accuracy.\cite{Kern1996,Kern1997}   
The surface geometries were obtained using the PBE functional, further details are given in Supplementary Sec. S4.\\
RPA formation energies of diamond surfaces are collected in Table \ref{tab:surfaces}. (100) and (111)-1db surfaces were included in training, whereas (110) and (111)-3db surfaces are left as independent tests for ML-RPA. For all surface terminations studied, ML-RPA correctly predicts the most stable surface orientation (underlined values). For example, of the clean surfaces (as cut), both RPA and ML-RPA predict the order (110) $<$ (111)-1db $<$ (100) $<$ (111)-3db, which can be qualitatively understood via coordination of the surface atoms.\cite{Kern1996} Note that the (111)-3db surface becomes increasingly competitive with the (111)-1db surface as more dangling bonds are eliminated via reconstruction (-1 db) and chemisorption (-2 db for H, -3 db for O). 

Including calculations of metastable surfaces (see Supplementary Sec. S4), we have assembled a database of 28 RPA surface formation energies in total. It is interesting to use these data as benchmark for other exchange-correlation functionals. Table \ref{tab:surface benchmark} shows that all functionals including ML-RPA tend to predict smaller formation energies than exact RPA (negative mean relative errors). In terms of accuracy, ML-RPA performs very well, with a mean absolute error of 70 meV per surface atom. Surface energies predicted by the meta-GGA functional SCAN as well as the vdW functionals PBE+TS and SCAN+rVV10 are also in very good agreement with exact RPA, with mean absolute errors slightly larger than ML-RPA.
Finally, we comment again on the stability of electronic self-consistancy. Going from ML-RPA@PBE to self-consistent ML-RPA, individual surface energies change by 60 meV per surface atom or less, and the mean absolute error of self-consistent ML-RPA (compared to exact RPA@PBE) is only 80 meV per surface atom. Thus, the accuracy of ML-RPA is not significantly diminished by electronic self-consistency.
\begin{table*}[!htb]  
\caption{RPA energies of formation for diamond surfaces, calculated via Eq. \eqref{eq:formation-surface} (in eV per surface atom). The most stable surface orientation for a given surface termination is underlined, and surfaces included in the ML-RPA training set are marked by asterisks. Surface calculations are performed in $2\times1$ supercells, with geometries illustrated in Fig. \ref{fig:surfaces}.}\label{tab:surfaces}
\begin{tabular}{lrrrrrrrr}  
\hline\hline\\\\[-4.\medskipamount]
&\multicolumn{2}{c}{$(100)^*$} & \multicolumn{2}{c}{(110)}  & \multicolumn{2}{c}{(111)-1db$^*$} & \multicolumn{2}{c}{(111)-3db}\\
\hline \\\\[-4.\medskipamount]
& exact & ML & exact & ML & exact & ML & exact & ML \\
\cline{2-3} \cline{4-5} \cline{6-7} \cline{8-9} \\\\[-4.\medskipamount]
clean         & 3.66              &  3.57            & \underline{2.09} & \underline{1.95} & 2.69               &  2.52              & 4.37  &  4.37 \\
reconstructed & {$2.08$}          &  2.03            & 1.70             & 1.59             & \underline{1.41}   &  \underline{1.34}  & 2.70  &  2.56 \\
+H (1 ML)     & {$0.20$}          &  0.11            & -0.11            & -0.12            & \underline{-0.19}  &  \underline{-0.24} & 0.17  &  0.11 \\
+O (1 ML)     &  \underline{1.99} &  \underline{1.91}&  3.23            & 3.21             &  {$2.99$}          &  2.94             & 2.90  &  2.79 \\
\hline \\\\[-4.\medskipamount]
\multicolumn{4}{l}{{$^*$included in the ML-RPA training set}} \\
\hline\hline
\end{tabular}
\end{table*}
\begin{table}[!htb]  
\caption{Surface formation energies from different exchange-correlation functionals compared to values from exact RPA. Mean signed error (MSE), mean absolute error (MAE) and maximum absolute error (MAX) are given in eV per surface atom. Averages are over 28 diamond surfaces of which 16 are in the ML-RPA training set, see Supplementary Section D.}\label{tab:surface benchmark}
\begin{tabular}{l @{\qquad} r @{\qquad} r @{\qquad} r}  
\hline\hline \\\\[-4.\medskipamount]
& MSE & MAE & MAX  \\
\hline \\\\[-4.\medskipamount]
LDA     &  -0.06 & 0.14   & 0.34   \\
PBEsol  & -0.10 & 0.12  & 0.33    \\
PBE      & -0.19 & 0.19  & 0.39   \\
RPBE     & -0.23 & 0.24  & 0.44   \\
\hline \\\\[-4.\medskipamount]
PBE+TS  & -0.01 & 0.08 & 0.25 \\
RPBE+D3(BJ) & -0.09 & 0.12 & 0.48 \\
optB88-vdW & -0.19 & 0.19 & 0.62 \\
rev-vdW-DF2 & -0.13  & 0.14  & 0.52 \\
rVV10           &  -0.18 & 0.18 & 0.57 \\
\hline \\\\[-4.\medskipamount]
SCAN & -0.08 & 0.11 & 0.25 \\
SCAN+rVV10 &-0.03 & 0.10 & 0.24 \\ 
\hline \\\\[-4.\medskipamount]
ML-RPA & -0.07 &  0.07  & 0.18   \\
\hline\hline
\end{tabular}
\end{table}


\subsection{Liquid water}

Water with its many anomalies is both an important and challenging system for first-principles molecular dynamics simulations.\cite{Brini2017} The role of the exchange-correlation functional for the description of liquid water and ice has been studied extensively.\cite{Macher2014,DelBen2015,Gillan2016,Chen2017} This has made water an interesting target for several recent MLFF\cite{Morawietz2016,Dasgupta2021,Yao2021} and ML-DFT\cite{Margraf2021,Dick2021} approaches. \\
In particular, \myciteauthor{Yao2021}\cite{Yao2021} used MLFF to perform RPA-level calculations for liquid water with the inclusion of nuclear quantum effects. They showed that the RPA can well reproduce experimental data for numerous water properties at different temperatures. Due to the small mass of the hydrogen atom, the oxygen-hydrogen radial distribution function (RDF) and especially the hydrogen-hydrogen RDF are significantly altered by nuclear quantum effects. Namely, classical MD predicts an oxygen-hydrogen RDF that is overstructured compared to experimental data, and even more so for the hydrogen-hydrogen RDF. In contrast, the oxygen-oxygen RDF is far less affected, especially for higher temperatures. 

To perform calculations on liquid water, we add 32 liquid water structures containing 8 water molecules to the ML-RPA training set, as well as 39 structures using larger supercells (31 or 32 water molecules). The structures are again sampled from MD trajectories created by earlier ML-RPA versions. In addition, the ML-RPA training set contains 6 structures sampling the water monomer. 

Accurate determination of the water RDFs require long MD simulations that are computationally expensive even without nuclear quantum corrections. Thus, we perform classical MDs using 64 water molecules and speed them up by combining ML-RPA with MLFF. That is, we train a machine learning force field ``on-the-fly''\cite{Jinnouchi2019,Jinnouchi2019a} using the energies and forces predicted by ML-RPA. In order to obtain an RPA reference for the radial distribution function, we also train an MLFF directly on exact RPA energies and forces (RPA-MLFF). The RPA-MLFF training set contains all water structures from the ML-RPA training set plus 30 additional structures containing 32 molecules. To validate our machine learning force fields, we also trained on-the-fly MLFFs for the vdW functionals RPBE+D3(BJ)\cite{Grimme2010,Grimme2011} and PBE+TS.\cite{Tkatchenko2009} Further details of the MLFF setups are given in Supplementary Sec. S5. 


\begin{figure}[!htb] \label{water_RDF}
\centering
\subfloat{\includegraphics [angle=0,width=0.55\linewidth,keepaspectratio=true] {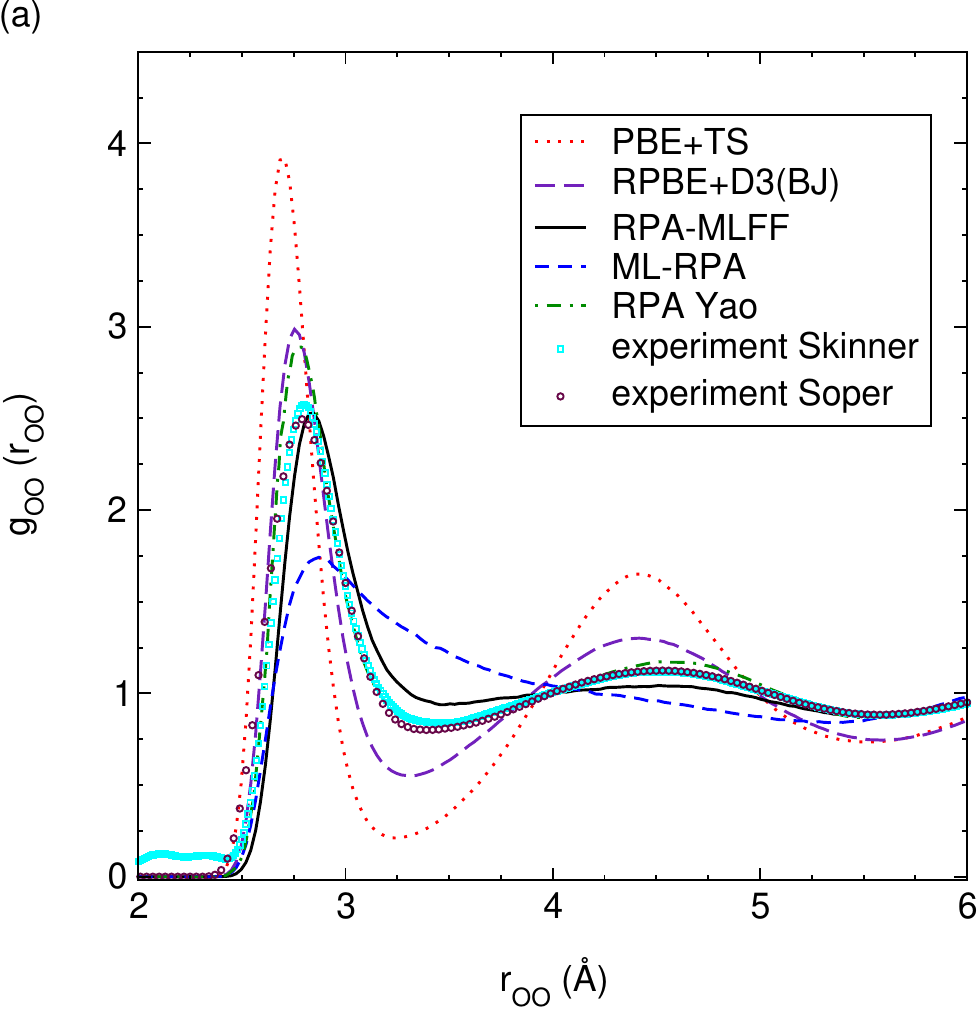}\label{fig:water_RDF_OO}}
\par\medskip
\subfloat{\includegraphics [angle=0,width=0.55\linewidth,keepaspectratio=true] {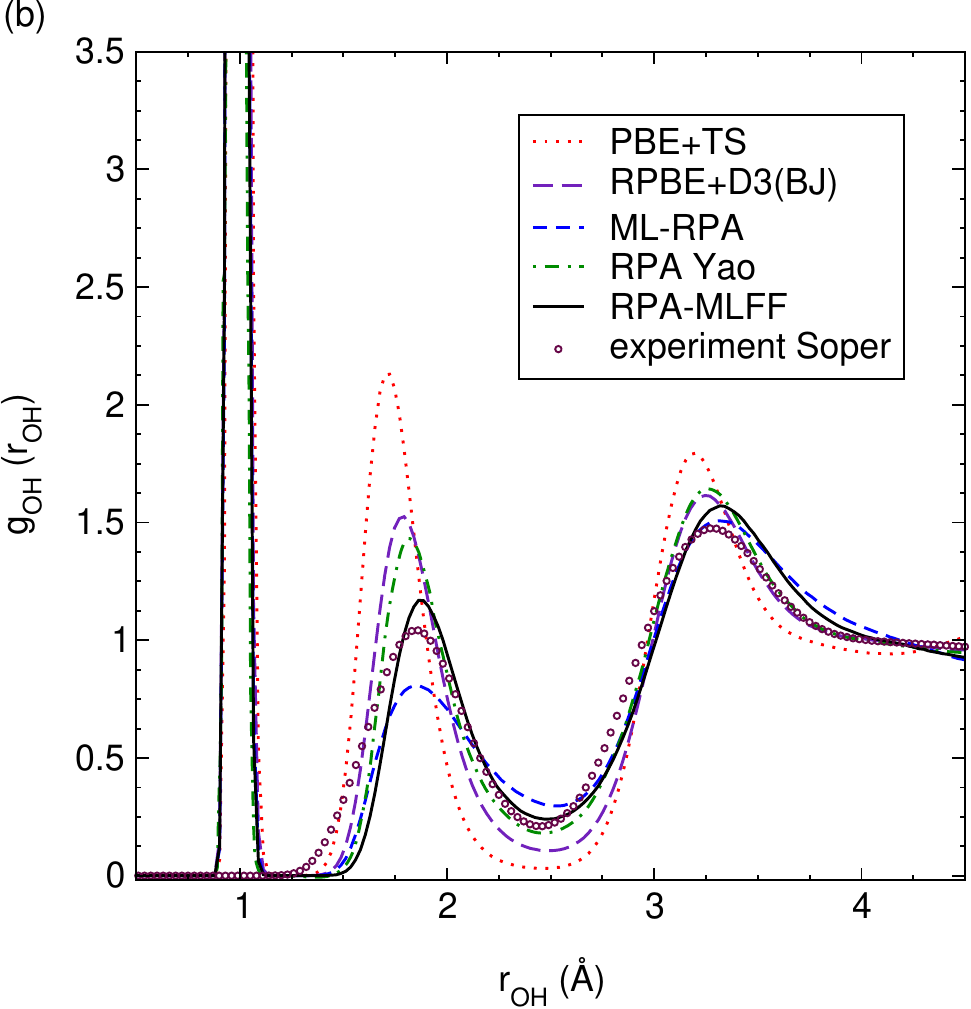}\label{fig:water_RDF_OH}}
\caption{Partial radial distribution functions of liquid water at ambient conditions ($\text{T}= 300 \text{ K}$, $\rho = 1 \text{ g/cm}^3$), details of the MD simulations are given in Supplementary Sec. S5. (a) oxygen-oxygen radial distribution function, (b) oxygen-hydrogen radial distribution function. RPA results from Yao and Kanai are extracted from Ref. \protect\onlinecite{Yao2021}, experimental data are taken from Refs. \protect\onlinecite{Skinner2013,Soper2013}. See text for a discussion of nuclear quantum effects.}
\label{fig:water_RDF}
\end{figure}

Fig \ref{fig:water_RDF} shows the oxygen-oxygen RDF, $g_{\rm OO}$, and oxygen-hydrogen RDF, $g_{\rm OH}$, at ambient conditions. 
First, we note that the RDFs obtained from PBE+TS and RPBE+D3(BJ) are in good agreement with respective literature results\cite{Zheng2018, Sakong2016} (for PBE+TS, we compared the RDF at T = 330 K, not shown). The PBE+TS RDF is clearly overstructured compared to experimental data. The water structure is ``too tetragonal'', even more so when the TS vdW correction is not included, see Ref. \onlinecite{Zheng2018}. The RPBE+D3(BJ) RDF is in better agreement with experiment, but is still somewhat overstructured. This is a specific effect of the Becke-Johnson (BJ) damping, that is, the RPBE+D3(0) RDF (using zero damping) is closer to experiment (see Ref. \onlinecite{Sakong2016}).
Next, the RDFs predicted by RPA-MLFF are less structured than the RPA references of \myciteauthor{Yao2021}, compare the first peaks and minima of both $g_{\rm OO}$ and $g_{\rm OH}$. This discrepancy is possibly due to technical convergence of either RPA calculation, in particular, basis set incompleteness errors. 
The fact that RPA-MLFF closely reproduces the first experimental peak height of $g_{\rm OO}$ is arguably accidental, since nuclear quantum effects are not included. For $g_{\rm OH}$, however, the result of \myciteauthor{Yao2021} is still more structured than experiment even with nuclear quantum effects included (see Fig. 1 in Ref. \onlinecite{Yao2021}), indicating that our RPA reference is potentially more accurate. 

That said, overall both RPA reference results agree well with experimental data. Turning to the water structure predicted by ML-RPA, there are some clear discrepancies from both RPA references. The first peak of $g_{\rm OO}$ is smaller in height, and the second peak is completely missing. The $g_{\rm OH}$ predicted by ML-RPA is also somewhat less structured than the RPA references, but the overall agreement is better than for $g_{\rm OO}$. Importantly, the discrepancies are mainly due to the actual ML-RPA density functional, since the on-the-fly MLFF is very accurate (see Supplementary Sec. S5). Specifically, atomic forces for all MLFFs trained here exhibit an root mean square error of roughly $30 \text{ meV \AA}^{-1}$, whereas the respective force errors for liquid water due to ML-RPA are roughly $90 \text{ meV \AA}^{-1}$.
It is also important to point out that the RPA-MLFF is a single-purpose force field, trained specifically to simulate liquid water in the bulk (and the water monomer). In contrast, the ML-RPA functional is not limited in this way as will be demonstrated in the following section.

\clearpage

\subsection{Smaller water clusters}
\begin{table}[!htb]  
\caption{Ground state properties of the $\text{H}_2\text{O}$ dimer. Dimer binding energies, $E_{\rm b}^{\rm dim}=-E[(\text{H}_2\text{O})_2]+2E[\text{H}_2\text{O}]$, are given in eV. Equilibrium bond lengths $d_{\rm OO}^{\rm dim}$ and $d_{\rm OH}^{\rm dim}$ are given in \AA. All calculations are performed fully self-consistently.} \label{tab:water dimer}
\begin{tabular}{l @{\qquad} r @{\qquad} r @{\qquad} r}
\hline\hline \\\\[-4.\medskipamount]
& $E_{\rm b}^{\rm dim}$ & $d_{\rm OO}^{\rm dim}$ & $d_{\rm OH}^{\rm dim}$  \\
\hline \\\\[-4.\medskipamount]
LDA       & 0.39  & 2.72   & 1.73   \\
PBEsol    & 0.27 & 2.80 &  1.82  \\
PBE        & 0.23  & 2.88 & 1.91   \\
RPBE      & 0.17 &  3.02  & 2.04   \\
\hline \\\\[-4.\medskipamount]
PBE+TS          &  0.24 & 2.89 & 1.92 \\
RPBE+D3(BJ)       & 0.21 & 2.94 & 1.97 \\
optB88-vdW   &   0.22  & 2.90 & 1.93 \\
rev-vdW-DF2 & 0.23 & 2.89 & 1.92 \\
rVV10          & 0.25 & 2.88 & 1.90 \\
\hline \\\\[-4.\medskipamount]
SCAN  & 0.25 & 2.84 & 1.88 \\
SCAN+rVV10 & 0.26 & 2.84 & 1.88 \\
\hline \\\\[-4.\medskipamount]
ML-RPA & 0.18 & 3.07 & 2.10 \\
RPA-MLFF & 0.21 & 3.28 & 2.30 \\
\hline \\\\[-4.\medskipamount]
CCSD(T)\cite{Lane2012} & 0.22 &  2.91  & 1.96 \\
\hline\hline
\end{tabular}
\end{table}
\begin{figure}[!htb] \label{hexamers_main}
\centering
\includegraphics [width=0.75\linewidth,keepaspectratio=true] {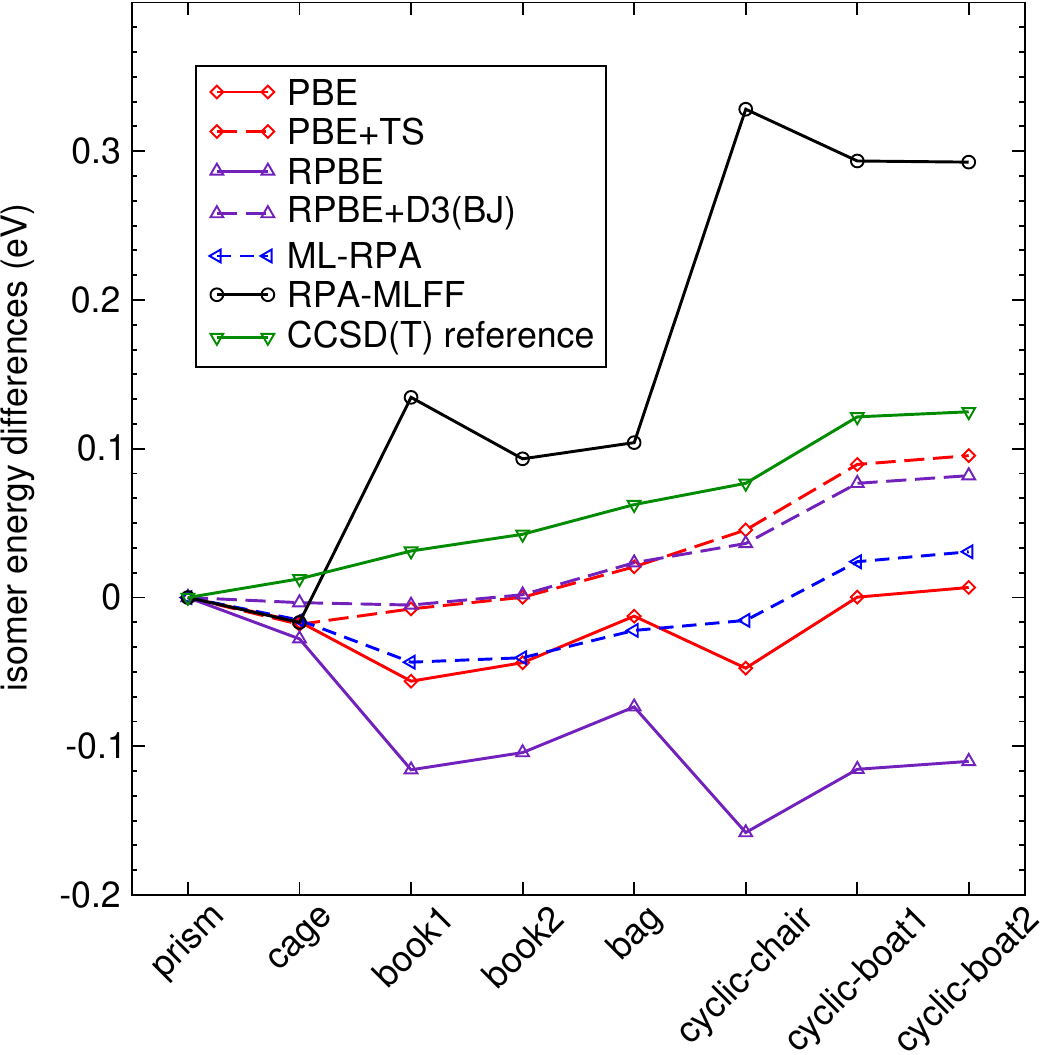}
\caption{Binding energy differences of eight $\text{H}_2\text{O}$ hexamers. Geometries and CCSD(T) reference data are taken from Ref. \protect\onlinecite{Reddy2016}. The vdW functionals PS+TS and RPBE+D3(BJ) largely correct the errors of their respective GGA base functionals, see also Supplementary Sec. S6. Lines drawn are only guides to the eye.}
\label{fig:hexamers_main}
\end{figure}

An important $\text{H}_2\text{O}$ benchmark is the performance for smaller water clusters.\cite{Gillan2016} Already the simple $\text{H}_2\text{O}$ dimer gives a predictive measure of the strength of a hydrogen bond in liquid water. Table \ref{tab:water dimer} shows that ML-RPA predicts a somewhat underbound dimer, consistent with the understructured liquid.
Going to larger clusters, the water hexamers present a difficult challenge for DFT, as three-dimensional structures (``prism'',''cage'',''bag'') compete energetically with two-dimensional ones (``book'',''chair'',''boat''). 
Fig. \ref{fig:hexamers_main} shows that ML-RPA erroneously predicts two-dimensional structures to be most stable, as do all GGA functionals (see also Supplementary Sec. S6). LDA and the SCAN meta-GGA functional perform well in this regard, but overbind the dimer, compare Table \ref{tab:water dimer}. All vdW functionals tested give excellent results for both the dimer as well as the hexamers. This confirms the critical role that vdW interactions play for the structure of water.\cite{Gillan2016,Morawietz2016} Further, RPA-MLFF clearly fails for water clusters, but we reiterate that it has been trained only for liquid water in the bulk. The fact that RPA-MLFF extrapolates is substantiated by large Bayesian error predictions: compared to typical bulk water configurations, the maximum force Bayesian error is three times as large for the hexamers and five times as large for the dimer.

Finally, we investigated cubic ice, specifically, the $\rm I_c$(a) proton-ordered ice phase as described in Ref. \onlinecite{Macher2014}. The equilibrium volume predicted by RPA-MLFF is $\text{32.3 \AA}^3{\text{ per H}}_2{\text O}$, in good agreement with the $\text{32.6 \AA}^3{\text{ per H}}_2{\text O}$ obtained using exact RPA. The ML-RPA equilibrium volume is somewhat smaller ($\text{31.2 \AA}^3{\text{ per H}}_2{\text O}$). For comparison, the equilibrium volumes predicted by PBE and PBE+TS are $\text{30.2 \AA}^3{\text{ per H}}_2{\text O}$ and $\text{30.0 \AA}^3{\text{ per H}}_2{\text O}$, respectively. In summation, ML-RPA provides a consistent but somewhat inaccurate description of water and ice. The fact that ML-RPA misses the second $g_{\rm OO}$ maximum, underbinds the water dimer, and provides a PBE-like description of the water hexamers and cubic ice strongly indicates that the current ML-RPA misses some crucial non-local interactions.
This is likely connected to the rather small cutoff radius used here for ML-RPA ($R_{\rm cut}=1.5 \text{ \AA}$). However, increasing the cutoff is not beneficial for the current ML-RPA, our tests indicate that this diminishes ML-RPA accuracy (not shown).


\subsection{Homogeneous electron gas}

\begin{figure}[!htb] \label{HEG}
\centering
\includegraphics [width=0.75\linewidth,keepaspectratio=true] {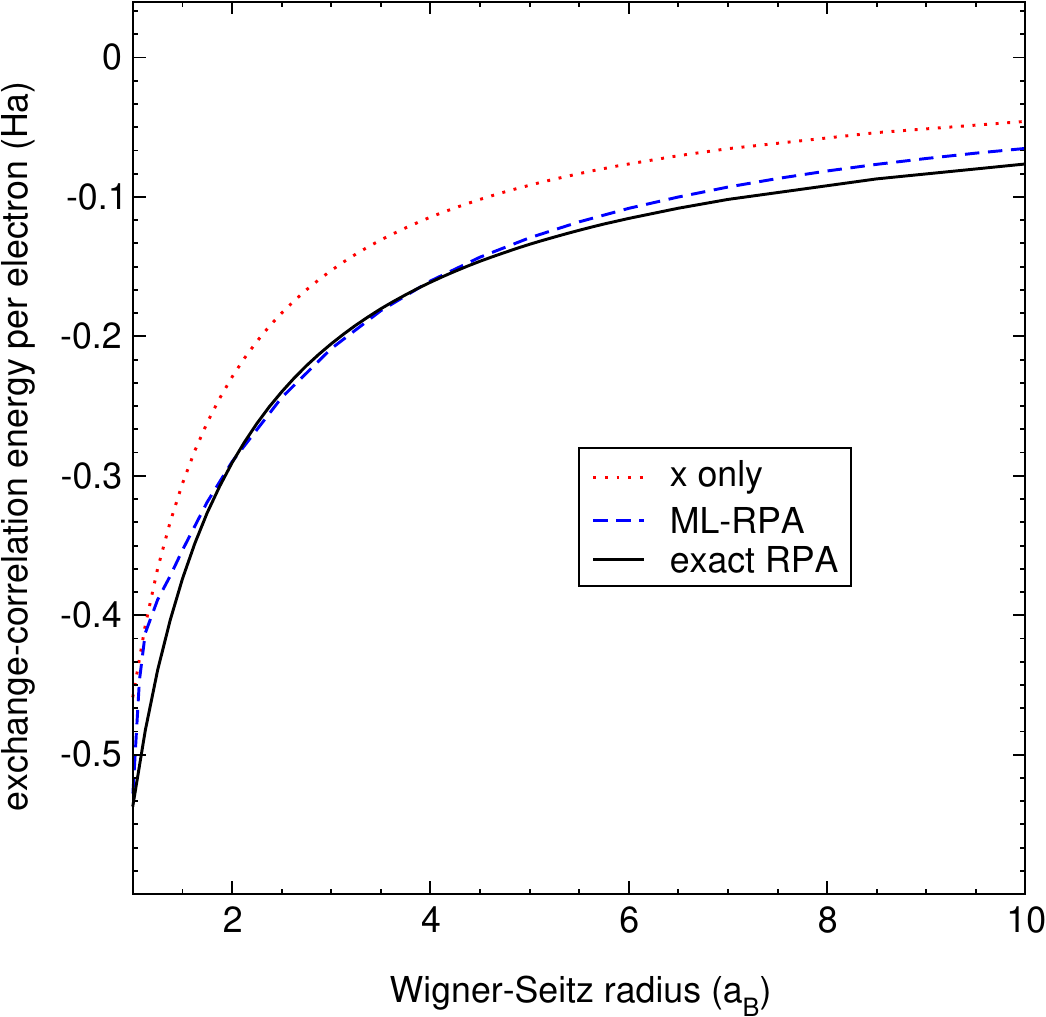}
\caption{Application of ML-RPA to the homogeneous electron gas. 
Exact RPA results were taken from Ref. \protect\onlinecite{Riemelmoser2020}, the exchange only approximation $\varepsilon_{\rm x, HEG}$ is also shown for comparison. No HEG data were added to the ML-RPA training set.}
\label{fig:HEG}
\end{figure}
\noindent To challenge the extrapolation abilities of our ML-RPA functional, we apply it to the homogeneous electron gas (HEG). The HEG constitutes an ``appropriate norm'', that is an important theoretical limit that density functionals should fulfill.  Due to symmetry, the HEG is completely characterized by the electron density $n$, or equivalently the Wigner-Seitz radius $r_{\rm s}=(3/4\pi n)^{1/3}$. Starting with LDA, many successful exchange-correlation functionals are designed such that they describe the HEG exactly, but this is not the case for the present ML-RPA functional. It is important to point out that the ground truth here is the exchange-correlation energy per electron as given by exact RPA. As the RPA is itself an approximation, $\varepsilon_{\rm xc, HEG}^{\rm RPA}$ differs from the usual LDA which is based on exact numerical data from Quantum Monte Carlo calculations.\cite{Ceperley1980}
Figure \ref{fig:HEG} shows that ML-RPA for intermediate densities ($r_{\rm s} \sim 2-5$) closely follows the exact RPA reference. Since this is the range of physical densities, ML-RPA seems to have learned the HEG indirectly from diamond and water data. Still, this excellent agreement comes somewhat surprising considering that no HEG data were explicitly added to the ML-RPA training set.  Furthermore, for small densities ($r_{\rm s} \gtrsim 5$), ML-RPA is slightly less accurate but still well behaved. Finally, for large densities ($r_{\rm s}\lesssim 2$), ML-RPA develops an unphysical kink. This indicates the presence of an ``extrapolation hole'', that is, the complete lack of training data can cause erratic behavior. 
\section{Discussion}\label{sec:Riemelmoser2023_5}

The LDA and different GGA functionals differ in the strength of their respective enhancement factors. This results, for example, in the following trend for cohesive energies of solids 
\begin{equation}
\text{LDA} > \text{PBEsol} > \text{PBE} > \text{RPBE} . 
\end{equation}
The same trend is also manifest for surface formation energies and molecular adsorption energies.\cite{Schimka2010} As different enhancement strengths fit better for different physical properties, this leads to a well-known trade-off for GGA functionals. This trade-off is visible also in the current study:
LDA and PBEsol give better diamond surface energies (Table \ref{tab:surface benchmark}). PBE and RPBE perform better for the water dimer (Table \ref{tab:water dimer}), but the trend is reversed for the hexamer puzzle (Supplementary Sec. S6). This means that no GGA functional can give a completely satisfactory description of liquid water and ice, see also Ref. \onlinecite{Gillan2016} for a more in-depth discussion.  We have demonstrated that the non-local gradient approximation used in our ML-RPA model can overcome the GGA trade-off to some extend. While ML-RPA does not exceed a GGA-level description of liquid water, it clearly outperforms all GGA functionals for the diamond surface benchmark.

Traditional routes beyond GGA are metaGGA and hybrid functionals on the one hand, and vdW functionals on the other hand. The RPA itself can be considered as the ``gold standard'' of vdW functionals.\cite{Klimes2012}  
Different vdW functionals tested here generally outperform their respective semi-local DFT base functionals, though the improvement is not always consistent. For example, the rVV10 vdW correction slightly increases the surface energies of the pristine SCAN functional, achieving good agreement with exact RPA (Table \ref{tab:surface benchmark}, Ref. \onlinecite{Patra2017}). On the other hand, SCAN already slightly overbinds the water dimer and gives a somewhat overstructured liquid. Thus, it is plausible that additional binding in the form of rVV10 can only deteriorate the performance of SCAN for water (Supplementary Sec. S6, Ref. \onlinecite{Wiktor2017}).
We reason that the rather small cutoff radius adapted here for ML-RPA is not sufficient to capture the full non-locality that is required for a complete description of liquid water (ML-RPA uses a cutoff radius of 1.5 \AA, whereas RPA-MLFF uses a larger cutoff of 6.0 \AA). However, simply increasing the cutoff radius is not an option with the current implementation and training database. We repeat that our tests show that a larger cutoff would diminish fit accuracy. Future work could instead focus on the construction of explicitly long-range descriptors as present in non-local vdW functionals or in recent extensions of MLFF frameworks.\cite{Grisafi2019}

Finally, a common problem of ML techniques is extrapolation, that is, one can only expect good performance if applications are similar enough to the respective training sets. Here, this point was demonstrated by the poor performance of RPA-MLFF for water clusters (RPA-MLFF has been trained only for bulk water and the water monomer). In contrast, ML-RPA gives consistent if inaccurate predictions, though it has less water training data. We speculate that this is due to ML-RPA descriptors being much more compact, which makes extrapolation more manageable. Specifically, here we used 8 density descriptors for ML-RPA versus 408 atomic descriptors for RPA-MLFF (204 for both O and H). It is worth pointing out that this difference will be exacerbated when more chemical species have to be described at the same time. That is, the number of descriptors in MLFF schemes generally scales unfavorably with the number of chemical species, whereas the density descriptors used for ML-RPA do not depend explicitly on atom type. On the other side, RPA-MLFF is in the present implementation undeniable superior in terms of raw fit accuracy. ML-RPA and RPA-MLFF thus strike different balances between model flexibility and universality.

When pushed even further outside of its training set, however, ML-RPA eventually extrapolates as well as demonstrated for the homogeneous electron gas. The obvious but costly solution to the extrapolation problem is the construction of ever larger training databases. A more elegant alternative would be to enforce ML-RPA to obey known exact constraints. Exact constraints have been used in the past to construct successful functionals such as the SCAN functional (``Strongly constrained and appropriately normed'').\cite{Sun2015} Recently, there have been some promising efforts to incorporate exact constraints also into machine learned density functionals.\cite{Hollingsworth2018,Dick2021,Nagai2022,Bystrom2022}

\section{Conclusion and Outlook}\label{sec:Riemelmoser2023_6}

In this work we have machine learned a substitute density functional based on the random-phase approximation. The ingredients of ML-RPA are density descriptors constructed analogously to the two- and three-body descriptors used for machine learned force fields. These ingredients can be considered as non-local extensions of the local density and its gradient. As a first application, we have constructed an ML-RPA functional for diamond and liquid water. We have demonstrated how such a functional can be used to enable RPA calculations at a larger scale. For a data set of 28 diamond surfaces, ML-RPA surpasses all tested GGA functionals in terms of accuracy and reaches the level of state-of-the-art vdW functionals. For liquid water, ML-RPA is less accurate and falls back to a GGA-level description, which we traced back to an insufficient description of non-local interactions.

Our ML-RPA scheme was demonstrated to learn fairly quickly from small amounts of exact RPA data, with the entire data base consisting of less than 200 structures. We credit this data efficiency to the inclusion of derivative information in the form of RPA exchange-correlation potentials, which are obtained via the optimized effective potential method. This is in close analogy to fitting atomic forces in MLFF. Generally, the tasks of machine learning atomic force fields and machine learning density functionals are closely related. We hope to see continued exchange of concepts and techniques, as we believe that both fields can benefit immensely from such a ``cross-fertilization'' of  ideas. 

Finally, our machine learning method using optimized effective potentials is general and not limited to the random-phase approximation. For example, beyond RPA theories can be constructed by including vertex correction to the screened Coulomb interaction and/or self energy. One can also envision to extract high accuracy exchange-correlation potentials from accurate coupled-cluster densities via Kohn-Sham inversion. Our method can also be applied to learn hybrid functionals, where large databases can be obtained more easily, thus facilitating large-scale hybrid functional simulations.

\begin{acknowledgement}

\noindent Computation time at the Vienna Scientific Cluster (VSC) is gratefully acknowledged. The authors thank K. Burke as well as K. Bystrom and B. Kozinsky for fruitful discussions. 

\end{acknowledgement}

\section*{Author declarations}

\subsection*{Conflict of interest}

\noindent The authors declare no competing interests.

\section*{Data availability}

The ML-RPA training data is freely available at \url{https://doi.org/10.25365/phaidra.418}.

\begin{suppinfo}

Analytical proof of Eqs. \eqref{eq:2-body limit} and \eqref{eq:3-body limit} (S1), analytical expressions for the ML-RPA exchange-correlation potentials (S2), details of our data sparsification scheme (S3), training details for ML-RPA  (S4) and RPA-MLFF (S5), water hexamer benchmark for more DFT functionals (S6).

\end{suppinfo}



\bibliography{Riemelmoser2023}

\end{document}


\section{Non-local density descriptors}\label{App:Riemelmoser2023_A}

In the following, we describe our density descriptors in more detail and comment on how they can be evaluated efficiently via the use of fast Fourier transforms (FFTs). 
We begin by expanding the electronic density around each grid point $\textbf{r}$ [Eq. (2) in the main text]
\begin{equation}
\begin{aligned}
n(\textbf{r}+\textbf{r}') f_{\rm cut}(r') = \sum_{nlm} c_{nlm}(\textbf{r}) \phi_{nl}(r')  Y_l^m(\widehat{\textbf{r}'}) .
\end{aligned}
\end{equation}
For the cutoff function $f_{\rm cut}$, we use the cosine cutoff proposed by \myciteauthor{Behler2007},\cite{Behler2007}
\begin{equation}
f(r') = \begin{cases}\frac{1}{2}\left[1+\cos\left(\frac{\pi r'}{R_{\rm cut}}\right)\right] & 0\leq r' \leq R_{\rm cut} \\
      0 & r\geq R_{\rm cut} . \end{cases} 
\end{equation}
For radial basis functions $\phi_{nl}$, we use spherical Bessel functions as in Ref. \onlinecite{Jinnouchi2019a},
\begin{equation}
\phi_{nl}(r') = j_l(q_{nl}r') ,
\end{equation}
where the $q_{nl}$ are chosen such that the basis functions vanish at $r'=R_{\rm cut}$. Such spherical Bessel functions form a complete basis on the interval $[0,R_{\rm cut}]$ and fulfill orthogonality relations of the kind
\begin{equation} \label{eq:orthogonality}
\begin{aligned}
&\int_0^{R_{\rm cut}} \text{d}r \; r^2 j_l(q_{nl}r)j_l(q_{n'l}r) = \delta_{nn'}A_{nl} .\\
\end{aligned}
\end{equation}
For example, in the case of $l=0$ one obtains
\begin{equation}
\phi_{n0}(r') = j_{0}(q_{n0}r') = \frac{\sin(q_{n0}r')}{q_{n0}r'} ,
\end{equation}
and demanding that $\sin(q_{n0}R_{\rm cut})=0$ yields the simple expressions
\begin{equation}
\begin{aligned}
& q_{n0} = \frac{n\pi}{R_{\rm cut}} \\
& A_{n0} = \frac{R_{\rm cut}^3}{2n^2\pi^2} \hspace{14pt} n=1,2,... 
\end{aligned}
\end{equation}
Next, we calculate the expansion coefficients $c_{nlm}$ using the orthogonality relations \eqref{eq:orthogonality}, as well as those for the $Y_l^m$,
\begin{equation}\label{eq:YLM orthogonality}
\int d\Omega \; Y_l^m(\hat{\textbf{r}}) Y_{l'}^{m'}(\hat{\textbf{r}}) = \delta_{ll'} \delta_{mm'} ,
\end{equation}
yielding
\begin{equation}\label{eq:cnlm real-space}
\begin{aligned}
c_{nlm}(\textbf{r}) &=  \int_0^{R_{\rm cut}} \text{d}r' {r'}^2 \frac{\phi_{nl}(r')f_{\rm cut}(r')}{A_{nl}}  \int \text{d}\Omega' Y^m_l (\widehat{\textbf{r}'}) n(\textbf{r}+\textbf{r}') \\
 &= \int_0^{R_{\rm cut}} \text{d}r' {r'}^2 \tilde{\phi}_{nl}(r')  \int \text{d}\Omega' Y^m_l (\widehat{\textbf{r}'}) n(\textbf{r}+\textbf{r}') ,
\end{aligned}
\end{equation}
where we have absorbed the coefficients $A_{nl}$ and the cutoff function $f_{\rm cut}$ to define modified basis functions $\tilde{\phi}_{nl}$,
\begin{equation}
\tilde{\phi}_{nl}(r') = \frac{\phi_{nl}(r')f_{\rm cut}(r')}{A_{nl}} .
\end{equation}
Next, we combine the Fourier representation of the density 
\begin{equation}\label{eq:density Fourier-space}
\begin{aligned}
n(\textbf{r}+\textbf{r}') &= \int \frac{\text{d}\textbf{q}}{(2\pi)^3}\; e^{i\textbf{q}(\textbf{r}+\textbf{r}')} n(\textbf{q}) ,
\end{aligned}
\end{equation}
and the plane-wave expansion,\cite{Arfken1999}
\begin{equation}
e^{i\textbf{q}\textbf{r}'} = 4\pi \sum_{lm} i^l j_l(qr') Y_l^m(\widehat{\textbf{q}})Y_l^m (\widehat{\textbf{r}'}) , 
\end{equation}
to rewrite Eq. \eqref{eq:cnlm real-space} as
\begin{equation}
\begin{aligned}
c_{nlm} (\textbf{r}) = 4\pi \int_0^{R_{\rm cut}} \text{d}r' {r'}^2 \tilde{\phi}_{nl}(r')  \int \text{d}\Omega' Y^m_l (\widehat{\textbf{r}'})  \int \frac{\text{d}\textbf{q}}{(2\pi)^3} e^{i\textbf{q}\textbf{r}} n(\textbf{q}) \sum_{l'm'} i^{l'} j_{l'}(qr') Y_{l'}^{m'}(\widehat{\textbf{q}})Y_{l'}^{m'} (\widehat{\textbf{r}'}) .
\end{aligned}
\end{equation}
To simplify the equation above, we use the orthogonality relations of the $Y_l^m$ [Eq. \eqref{eq:YLM orthogonality}] and define the spherical Bessel transform,  
\begin{equation}\label{eq:clmn Fourier-space}
\begin{aligned}
&\tilde{\phi}_{nl}(q) = 4\pi \int_0^{R_{\rm cut}} \text{d}r \; {r'}^2 j_{l}(qr') \tilde{\phi}_{ln}(r') \\
\leftrightarrow \; &\tilde{\phi}_{nl}(r') = \frac{1}{2\pi^2}\int \text{d}q \; q^2 j_l(qr') \tilde{\phi}_{nl}(q) ,
\end{aligned}
\end{equation}
to finally arrive at the concise expression
\begin{equation}
c_{nlm}(\textbf{r}) = i^l \int \frac{\text{d}\textbf{q}}{(2\pi)^3}  e^{i\textbf{q}\textbf{r}} n(\textbf{q}) \tilde{\phi}_{nl}(q) Y_l^m(\widehat{\textbf{q}}) .
\end{equation}
This equation is evaluated numerically on an real-space grid,
\begin{equation}
c_{nlm}(\textbf{r}) = i^l \sum_{\textbf{q}}  e^{i\textbf{q}\textbf{r}} n(\textbf{q}) \tilde{\phi}_{nl}(q) Y_l^m(\widehat{\textbf{q}}) .
\end{equation}
Thus, the coefficients $c_{nlm}(\textbf{r})$ can be calculated \textit{simultaneously for all real-space gridpoints} via Fourier transform, and FFT achieves $\mathcal{N_{\rm FFT}\ln N_{\rm FFT}}$ scaling with respect to the number of real-space grid points $N_{\rm FFT}$.
 
To form descriptors for ML, one seeks rotationally invariant combinations of the coefficients $c_{nlm}$, as discussed by Bartok \textit{et al.}\cite{Bartok2013} The $c_{n00}$ are already rotationally invariant, following from the fact that $Y_0^0=1/\sqrt{4\pi}$ is scalar. This gives us the two-body descriptors, $X^{(2)}_n=c_{n00}$. Further rotational invariants can be formed as follows\cite{Bartok2013,Jinnouchi2019a}
\begin{equation}
p_{nn'l} = \sqrt{\frac{8\pi^2}{2l+1}} \sum_{m} c_{nlm} c_{n'lm} .
\end{equation}
For $l=1$, these rotational invariants can be understood as all possible dot products between vectors $\textbf{c}_{n}=\{c_{n1x},c_{n1y},c_{n1z}\}$. That the $\textbf{c}_{n}$ are vectors follows from the vector property of $\textbf{Y}_1=\{Y_1^x,Y_1^y,Y_1^z\}$ (we use the real spherical harmonics, where this vector property is clearly manifest)
\begin{equation}\label{eq:Y1M}
\begin{aligned}
Y_1^x = \sqrt{\frac{3}{4\pi}} \frac{x}{\sqrt{x^2+y^2+z^2}} \\
Y_1^y = \sqrt{\frac{3}{4\pi}} \frac{y}{\sqrt{x^2+y^2+z^2}} \\
Y_1^z = \sqrt{\frac{3}{4\pi}} \frac{z}{\sqrt{x^2+y^2+z^2}} .
\end{aligned}
\end{equation}
To form three-body descriptors that have the same dimension as the $X^{(2)}_n$, we first form descriptors $X_{nn'l}^{(3)}$,
\begin{equation}\label{eq:off-diagonal descriptors}
X_{nn'l}^{(3)} = \frac{\sigma^{(3)}}{R_{\rm cut}}\frac{ \sum_{m} c_{nlm} c_{n'lm}}{(\sum_{m} c_{nlm}^2)^{1/4}(\sum_{m} c_{n'lm}^2)^{1/4}} .
\end{equation}
Next, we limit the $X_{nn'l}^{(3)}$ to $l=1$ and neglect off-diagonal elements $n\neq n'$, obtaining the three-body descriptors $X_n^{(3)}$ [compare Eq. (4) in the main text]
\begin{equation}\label{eq:diagonal descriptors}
\begin{aligned}
X_n^{(3)} &= X_{nn'l}^{(3)} \delta_{nn'} \delta_{l1} \\
&= \frac{\sigma^{(3)}}{R_{\rm cut}}\sqrt{c_{n1x}^2+c_{n1y}^2+c_{n1z}^2} .
\end{aligned}
\end{equation}
We have verified that this last approximation works well for the application to diamond and liquid water. Generally, however, the validity of the approximation depends on the concrete training set as well as on the radial basis functions used.\cite{Bartok2013} Here, we use four radial basis functions for both $l=0$ and $l=1$, thus we have in total $4+4=8$ density descriptors. Note that we use the weight factor $\sigma^{(3)}/R_{\rm cut}$ in Eq. \eqref{eq:off-diagonal descriptors}, rather than $({8\pi^2/3})^{1/4}$ which would correspond to $\sqrt{p_{nn1}}$. 
The possibility of using different weight factors for the three-body descriptors was also discussed in detail by \myciteauthor{Jinnouchi2020}\cite{Jinnouchi2020} 

\subsection{Approaching GGA in the small cutoff limit}

In Eqs. (3) and (5) in the main text, we have stated that in the limit of small cutoffs, the two-and three-body descriptors reduce to the local density and its gradient, respectively. A detailed proof of these limits is given in the following. 
For a small enough cutoff radius $R_{\rm cut}$, the density inside the cutoff sphere varies little, assuming that the density is smooth enough. That means one can perform the gradient expansion of the density\cite{Kohn1965} around $\textbf{r}$
\begin{equation}\label{eq:gradient expansion}
\begin{aligned}
&n(\textbf{r}+\textbf{r}') \approx n(\textbf{r}) +  r_\alpha' \nabla_\alpha n(\textbf{r}) + \frac{1}{2}r_\alpha' r_\beta' \nabla_\alpha \nabla_\beta n(\textbf{r}) + \frac{1}{6}r_\alpha' r_\beta' r_\gamma' \nabla_\alpha \nabla_\beta \nabla_\gamma n(\textbf{r}) + ... ,
\end{aligned}
\end{equation}
where $\alpha$, $\beta$ and $\gamma$ are Cartesian indices and we use the Einstein sum convention. The same limit can also be obtained for fixed cutoff radius by considering artificial weakly varying densities ($\nabla n \rightarrow 0$, in the extreme case, one obtains the homogeneous electron gas). In other words, the dimensionless expansion parameter is $r'\nabla n$, and that can be small when either $r'$ or $\nabla n$ is fixed and the other one goes to zero.

We insert the gradient expansion \eqref{eq:gradient expansion} into the expression for the coefficients $c_{nlm}$, Eq. \eqref{eq:cnlm real-space}, yielding
\begin{equation}\label{eq:cnlm approximation}
\begin{aligned}
c_{nlm} (\textbf{r}) &\approx \int_0^{R_{\rm cut}} \text{d}r'\; {r'}^2 \tilde{\phi}_{nl}(r')
 \int \text{d} \Omega'\; Y_l^m(\widehat{\textbf{r}'}) \\ &\times\left[n(\textbf{r}) + r'_\alpha\nabla_\alpha n(\textbf{r}) + \frac{1}{2}r_\alpha' r_\beta' \nabla_\alpha \nabla_\beta n(\textbf{r}) + \frac{1}{6}r_\alpha' r_\beta' r_\gamma' \nabla_\alpha \nabla_\beta \nabla_\gamma n(\textbf{r}) +...  \right] .
\end{aligned}
\end{equation}
When evaluating the integral above, the first order term in the gradient expansion vanishes for $l=0$. This is due to the anti-symmetry of the integrand. Likewise, the zeroth and second order terms vanish for $l=1$. Thus, for the $c_{n00}$ (two-body descriptors), Eq. \eqref{eq:cnlm approximation} simplifies to
\begin{equation}
\begin{aligned}
c_{n00} (\textbf{r}) &\approx \sqrt{4\pi} \int_0^{R_{\rm cut}} \text{d}r'\; {r'}^2 \tilde{\phi}_{n0}(r') n(\textbf{r}) + \frac{\sqrt{4\pi} }{6} \int_0^{R_{\rm cut}} \text{d}r'\; {r'}^4 \tilde{\phi}_{n0}(r') \nabla^2n (\textbf{r}) + ... ,
\end{aligned}
\end{equation}
where we have used
\begin{equation}\label{eq:angular integral}
\int \text{d}\Omega'\; r'_\alpha r'_\beta = \frac{4\pi}{3} {r'}^2 \delta_{\alpha\beta} .
\end{equation}
Further, the radial integrals give numerical constants, the first being independent of $R_{\rm cut}$, and the second being proportional to $R_{\rm cut}^2$ (this can be seen by performing a variable transform $x=r'/R_{\rm cut}$). The two-body descriptors in the limit of small cutoffs therefore have the expansion [compare Eq. (3) in the main text]
\begin{equation}
X^{(2)}_n (\textbf{r}) \propto n(\textbf{r}) + \text{const }\times R_{\rm cut}^2\nabla^2 n(\textbf{r}) + ...
\end{equation}
Next, we use Eq. \eqref{eq:cnlm approximation} to approximate the $l=1$ coefficients $c_{n1\alpha}$,
\begin{equation}
\begin{aligned}
c_{n1\alpha}(\textbf{r}) &\approx \int_0^{R_{\rm cut}} \text{d}r' \;{r'}^2 \tilde{\phi}_{n1}(r') \int \text{d}\Omega' \; Y_{1}^{\alpha}(\widehat{\textbf{r}'}) \left[r'_\beta \nabla_\beta n(\textbf{r})  + \frac{1}{6}r'_\beta r'_\gamma r'_\delta \nabla_\beta \nabla_\gamma \nabla_\delta n(\textbf{r})  + ...\right] \\
&= \sqrt{\frac{3}{4\pi}} \int_0^{R_{\rm cut}} \text{d}r'\; r' \tilde{\phi}_{n1}(r') \int d\Omega' \; r'_\alpha \left[r'_\beta \nabla_\beta n(\textbf{r})  + \frac{1}{6}r'_\beta r'_\gamma r'_\delta \nabla_\beta \nabla_\gamma \nabla_\delta n(\textbf{r})  + ...\right] ,
\end{aligned}
\end{equation}
where in the second line we have plugged in Eq. \eqref{eq:Y1M} for the $Y^\alpha_1$.  To simplify this expression, we use Eq. \eqref{eq:angular integral} for leading order term and the identity 
\begin{equation}
\int d\Omega' {r_\alpha'}{r_\beta'}{r'_\gamma}r'_\delta = \frac{4\pi}{15}{r'}^4 \left(\delta_{\alpha\beta} \delta_{\gamma\delta} + \delta_{\alpha\gamma}\delta_{\beta\delta} + \delta_{\alpha\delta}\delta_{\beta\gamma}\right) 
\end{equation}
for next-to-leading order term, yielding 
\begin{equation}
\begin{aligned}
&c_{n1\alpha}(\textbf{r}) \approx \\
& \sqrt{\frac{3}{4\pi}} \bigg\{ \left[\int_0^{R_{\rm cut}} \text{d}r' \; {r'}^3 \tilde{\phi}_{n1}(r')\right] \nabla_\alpha n(\textbf{r}) + \frac{1}{10}\left[\int_0^{R_{\rm cut}}\text{d}r' \; {r'}^5 \tilde{\phi}_{n1}(r')\right] \nabla_\alpha \nabla_\beta\nabla_\beta n(\textbf{r}) + ... \bigg\} \\
&\propto R_{\rm cut} \nabla_\alpha n (\textbf{r}) + \text{const}\times R_{\rm cut}^3 \nabla_\alpha\nabla_\beta\nabla_\beta n (\textbf{r}) + ...
\end{aligned}
\end{equation}
Next, we form the scalar products $c_{n1\alpha}c_{n'1\alpha}$ needed for the three-body descriptors. The leading order term in the expansion of these scalar products is proportional to the scalar product $\nabla_\alpha n(\textbf{r}) \nabla_\alpha n (\textbf{r}) = |\bm{\nabla} n(\textbf{r})|^2$, and the next-to-leading order term involves products of terms proportional to $\nabla_\alpha n(\textbf{r}) \nabla_\alpha \nabla_\beta \nabla_\beta n(\textbf{r})=\bm{\nabla} n(\textbf{r}) \cdot \bm{\nabla}\nabla^2 n(\textbf{r})$, thus
\begin{equation}
\begin{aligned}
&c_{n1\alpha}(\textbf{r})c_{n'1\alpha}(\textbf{r}) \propto R_{\rm cut}^2 |\bm{\nabla}n(\textbf{r})|^2 + \text{const}\times R_{\rm cut}^4\bm{\nabla}n(\textbf{r})\cdot \bm{\nabla}\nabla^2 n(\textbf{r})+ ...\\
\end{aligned}
\end{equation}
Inserting this result in the definition \eqref{eq:off-diagonal descriptors}, we finally obtain 
\begin{equation}
X_{nn'1}^{(3)} (\textbf{r}) \propto |\bm{\nabla}n(\textbf{r})| + \mathcal{O}(R_{\rm cut}^2) \hspace{14pt} \text{for } R_{\rm cut} \to 0 .
\end{equation}
As the three-body descriptors are simply the diagonal elements ($n=n'$) of the $X_{nn'1}$, this concludes the proof of the limit stated in Eq. (5) in the main text. Note that this limit extends also to the more general case of off-diagonal descriptors ($n\neq n'$).

\section{Machine learning DFT via optimized effective potentials}
\label{App:Riemelmoser2023_B}

In the following, we describe our ML scheme in more detail and discuss some challenges inherent to the use of optimized effective potentials (OEP). Further, we will derive analytic expression for the machine learned exchange-correlation potentials $v_{\rm xc}^{\rm ML-RPA}$ and show how they can be efficiently evaluated using FFTs.

We begin by briefly motivating our ML scheme via analogy to MLFF.
The starting point for MLFF schemes is the atomic density,
\begin{equation}
n_{\rm atom}(\textbf{r}) = \sum_{i}^{\rm atoms}  \delta(\textbf{r}-\textbf{R}_i) .
\end{equation}
This atomic density is usually smoothed by replacing the delta function above by a Gaussian. The central assumption in MLFF is that the total energy can by decomposed into a sum of atomic energies $\varepsilon_i$, which depend on two- and three-body descriptors (collected in a supervector $\textbf{X}_{\rm atom}$),
\begin{equation}\label{eq:atomic decomposition}
E = \sum_{i}^{\rm atoms} \varepsilon_i [\textbf{X}_{\rm atom}(\textbf{R}_i)] .
\end{equation}
In DFT, the central quantity is the electronic density $n$. We do not apply smearing to $n$, as the electronic density is already a smooth object. Analogously to Eq. \eqref{eq:atomic decomposition}, one can formulate the assumption that the exchange-correlation energy can be written as an integral of energy densities depending on two- and three-body descriptors (supervector $\textbf{X}$),
\begin{equation}
\begin{aligned}
E_{\rm xc} = \int \text{d} \textbf{r}\; n(\textbf{r}) \varepsilon_{\rm xc}[\textbf{X}(\textbf{r})] .
\end{aligned}
\end{equation}
We further pull out a factor $\varepsilon_{\rm x, HEG}(\textbf{r})$, such that the enhancement factor $F_{\rm xc}$ is approximated rather than $\varepsilon_{\rm xc}$. In other words, we use the LDA exchange as a baseline for ML-RPA, yielding the ansatz [Eq. (7) in the main text]
\begin{equation}\label{eq:ML-RPA ansatz copy}
E_{\rm xc}^{\rm ML-RPA} = \int \text{d} \textbf{r}\; n( \textbf{r}) \varepsilon_{\rm x, HEG}[n(\textbf{r})] F_{\rm xc}^{\rm ML-RPA}[\textbf{X} (\textbf{r})] .
\end{equation}
For the functional form of $F_{\rm xc}^{\rm ML-RPA}$, we use a Gaussian kernel [Eq. (8) in the main text]
\begin{equation}\label{eq:Gaussian kernel copy}
F_{\rm xc}^{\rm ML-RPA}[\textbf{X}(\textbf{r})] = \sum_{i_B} w_{i_B}\exp\left\{-\frac{[\textbf{X}(\textbf{r})-\textbf{X}^{i_B}]^2}{2\sigma^2}\right\},    
\end{equation}
where the kernel width $\sigma$ is an ML hyperparameter, the $\textbf{X}^{i_B}$ are representative control points and the $w_{i_B}$ the corresponding weights. Combing Eqs. \eqref{eq:ML-RPA ansatz copy} and \eqref{eq:Gaussian kernel copy}, we obtain 
\begin{equation}
\begin{aligned}
&E^{\rm ML-RPA}_{\rm xc} = \int \text{d} \textbf{r}\; n( \textbf{r}) \varepsilon_{\rm x, HEG}[n(\textbf{r})] \sum_{i_B} w_{i_B} \exp\left\{-\frac{[\textbf{X}(\textbf{r})-\textbf{X}^{i_B}]^2}{2\sigma^2}\right\} .
\end{aligned}
\end{equation}
Evaluating the functional derivative, $v_{\rm xc}^{\rm ML-RPA}=\delta E_{\rm xc}^{\rm ML-RPA}/\delta n(\textbf{r})$, yields a local term from the derivative of $n(\textbf{r})\varepsilon_{\rm x, HEG}(\textbf{r})$ and a non-local term which stems from the dependence of the descriptors $\textbf{X}(\textbf{r}')$ on the density $n(\textbf{r})$ 
\begin{equation}
\begin{aligned}
v_{\rm xc}^{\rm ML-RPA}(\textbf{r}) 
&= \int \text{d}\textbf{r}' \frac{\delta}{\delta n(\textbf{r})} \left\{n(\textbf{r}')\varepsilon_{\rm x, HEG} [n(\textbf{r}')]\right\} F_{\rm xc}^{\rm ML-RPA}(\textbf{r}') \\
&+ \int \text{d}\textbf{r}' n( \textbf{r}') \varepsilon_{\rm x, HEG}[n(\textbf{r}')] \frac{\delta}{\delta n(\textbf{r})} \left\{F_{\rm xc}^{\rm ML-RPA}[\textbf{X}(\textbf{r}')]\right\}  \\
&= v_{\rm xc, loc}^{\rm ML-RPA}(\textbf{r}) + v_{\rm xc, nl}^{\rm ML-RPA} .
\end{aligned}
\end{equation}
The local term is easily evaluated using $\delta n(\textbf{r}')/\delta n(\textbf{r})=\delta(\textbf{r}-\textbf{r}')$, yielding
\begin{equation}
\begin{aligned}
v_{\rm xc, loc}^{\rm ML-RPA}(\textbf{r}) &= \frac{4}{3} \varepsilon_{\rm x, HEG} [n(\textbf{r})] F_{\rm xc}^{\rm ML-RPA}[\textbf{X}(\textbf{r})]  \\
&= \frac{4}{3} \varepsilon_{\rm x, HEG} [n(\textbf{r})]  \sum_{i_B} w_{i_B} \exp\left\{-\frac{[\textbf{X}(\textbf{r})-\textbf{X}^{i_B}]^2}{2\sigma^2}\right\} .
\end{aligned}
\end{equation}
The non-local term is more complicated, but we will show in the following that FFT can be employed once again for its efficient evaluation.
Using Eq. \eqref{eq:Gaussian kernel copy} and applying the chain rule, we obtain
\begin{equation}\label{eq:ML-RPA vxc non-local}
\begin{aligned}
v_{\rm xc, nl}^{\rm ML-RPA}(\textbf{r}) &= \int \text{d}\textbf{r}' n( \textbf{r}') \varepsilon_{\rm x, HEG}[n(\textbf{r}')] \\
&\times \sum_{i_B} w_{i_B} \exp\left\{-\frac{[\textbf{X}(\textbf{r}')-\textbf{X}^{i_B}]^2}{2\sigma^2}\right\} \sum_{i}^{N_{\rm des}}\frac{-[X_i(\textbf{r}')-\textbf{X}^{i_B}]}{\sigma^2}\sum_{nml} \frac{\partial X_i(\textbf{r}')}{\partial c_{nlm}(\textbf{r}')}  \frac{\delta c_{nlm}(\textbf{r}')}{\delta n(\textbf{r})} .
\end{aligned}
\end{equation}
As the descriptors $X_i(\textbf{r}')$ depend on the expansion coefficients $c_{nlm}(\textbf{r}')$ in a simple algebraic fashion, the complicated non-locality is thus due to the last term only. Inserting the expressions \eqref{eq:clmn Fourier-space} and \eqref{eq:density Fourier-space} yields
\begin{equation}
\begin{aligned}
 \frac{\delta c_{nlm}(\textbf{r}')}{\delta n(\textbf{r})} &= \frac{\delta}{\delta n(\textbf{r})} i^l \int \frac{\text{d}\textbf{q}}{(2\pi)^3} e^{i\textbf{q}\textbf{r}'} \left[\int \text{d} \textbf{r}''\; e^{-i\textbf{q}\textbf{r}''} n(\textbf{r}'') \right] \tilde{\phi}_{nl}(q) Y_l^m(\widehat{\textbf{q}}) \\
 &=  i^l \int \frac{\text{d}\textbf{q}}{(2\pi)^3} \int \text{d} \textbf{r}''\;  e^{i\textbf{q}(\textbf{r}'-\textbf{r}'')} \delta(\textbf{r}-\textbf{r}'') \tilde{\phi}_{nl}(q) Y_l^m(\widehat{\textbf{q}}) \\
 &= i^l \int \frac{\text{d}\textbf{q}}{(2\pi)^3} e^{i\textbf{q}(\textbf{r}-\textbf{r}')} \tilde{\phi}_{nl}(q) Y_l^m(-\widehat{\textbf{q}}),
\end{aligned}
\end{equation}
where in the last line we have substituted $\textbf{q}\mapsto -\textbf{q}$. Next, we define the intermediate quantities $\eta_{nlm}$, which we evaluate numerically via FFT,
\begin{equation}\label{eq:etanlm}
\begin{aligned}
\eta_{nlm} (\textbf{q}) &= \frac{1}{N_{\rm FFT}} \sum_{\textbf{r}'} e^{-i\textbf{q}\textbf{r}'} n( \textbf{r}') \varepsilon_{\rm x, HEG}[n(\textbf{r}')] \\
&\times \sum_{i_B} w_{i_B} \exp\left\{\frac{-[\textbf{X}(\textbf{r}')-\textbf{X}^{i_B}]^2}{2\sigma^2}\right\} \sum_{i}^{N_{\rm des}}\frac{-[X_i(\textbf{r}')-\textbf{X}^{i_B}]}{\sigma^2}\frac{\partial X_i(\textbf{r}')}{\partial c_{nlm}(\textbf{r}')} .
\end{aligned}
\end{equation}
With the help of the $\eta_{nlm}$, we can rewrite Eq. \eqref{eq:ML-RPA vxc non-local} in the compact form
\begin{equation}\label{eq:ML-RPA vxc non-local efficient}
v_{\rm xc, nl}^{\rm ML-RPA}(\textbf{r}) = i^l \sum_{\textbf{q}} e^{i\textbf{q}\textbf{r}} \sum_{nlm} \eta_{nlm}(\textbf{q}) \tilde{\phi}_{nl}(q) Y_l^m(-\widehat{\textbf{q}}) ,
\end{equation}
which can be directly evaluated via FFT as well. In summary, Eqs. \eqref{eq:etanlm} and \eqref{eq:ML-RPA vxc non-local efficient} allow us to evaluate the ML-RPA exchange-correlation potential on all real-space grid points $\textbf{r}$ using a small number of FFTs. Thus, we have  \textit{avoided the evaluation of double integrals} by applying FFT throughout. Therefore, the overall computational cost of evaluating $v_{\rm xc}^{\rm ML-RPA}$ scales only as $\mathcal{O}(N_{\rm FFT}\ln N_{\rm FFT})$ rather than $\mathcal(N_{\rm FFT}^2)$ with respect to the number of real-space grid points $N_{\rm FFT}$. 

A potential pitfall in using the OEP method for ML applications is the fact that $v_{\rm xc}^{\rm RPA}(\textbf{r})$ is in practice determined only up to a constant shift. Inspired by the work of \myciteauthor{Nagai2018},\cite{Nagai2018} we circumvent this problem by defining auxiliary exchange-correlation potentials $\tilde{v}_{\rm xc}^{\rm RPA}$,
\begin{equation}\label{eq:potential shifts}
\tilde{v}_{\rm xc}^{\rm RPA}(\textbf{r}) = v_{\rm xc}^{\rm RPA}(\textbf{r}) + \frac{E_{\rm xc}^{\rm RPA}-\int \text{d}\textbf{r}'n(\textbf{r}')v_{\rm xc}^{\rm RPA}(\textbf{r}')}{\int \text{d}\textbf{r}'n(\textbf{r}')} .
\end{equation}
Thus, the $\tilde{v}_{\rm xc}^{\rm RPA}$ are shifted with respect to the $v_{\rm xc}^{\rm RPA}$ such that they integrate to $E_{\rm xc}^{\rm RPA}$,
\begin{equation}
\int \text{d} \textbf{r}\; n(\textbf{r}) \tilde{v}_{\rm xc}^{\rm RPA}(\textbf{r}) \overset{!}{=} E_{\rm xc}^{\rm RPA} .
\end{equation}
In fitting, we equate the auxiliary potentials with their ML-RPA analogs, $\tilde{v}_{\rm xc}^{\rm ML-RPA}$, 
\begin{equation}
\begin{aligned}
&\tilde{v}_{\rm xc}^{\rm ML-RPA}(\textbf{r}) = v_{\rm xc}^{\rm ML-RPA}(\textbf{r}) + \frac{E_{\rm xc}^{\rm ML-RPA}-\int \text{d}\textbf{r}' \; n(\textbf{r}')v_{\rm xc}^{\rm ML-RPA}(\textbf{r}')}{\int \text{d}\textbf{r}' \; n(\textbf{r}')} .
\end{aligned}
\end{equation}
Thus, any information regarding absolute values of the OEP potentials is circumvented. Similar shifted exchange-correlation potentials occur also in the ML scheme of Tozer \textit{et al.},\cite{Tozer1998} the Becke-Johnson method\cite{Becke2006} and the Levy-Zahariev formulation of DFT.\cite{Levy2014} Here, however, the auxiliary potentials are used only as intermediate quantities for fitting, and once an ML-RPA functional has been learned, standard exchange-correlation potentials $v_{\rm xc}^{\rm ML-RPA}$ are extracted for applications. 

To find the weights $w_{i_B}$, we fit to exchange-correlation energies and shifted exchange-correlation potentials at selected points $\textbf{r}_k$ for all structures $\alpha$ contained in the training set. We demand that ML-RPA reproduces the reference data in a least square sense and apply appropriate weights, yielding the loss function
\begin{equation}\label{eq:L2 loss}
\begin{aligned}
&\mathcal{L}_2 = \frac{1}{N_{\rm struct}} \sum_{\alpha}^{N_{\rm struct}} c_E \mathcal{L}_{2,E}^\alpha + (1-c_E) \mathcal{L}_{2,v}^\alpha \\
& \mathcal{L}_{2,E}^\alpha = \frac{1}{2} \frac{1}{N_e^\alpha} \left(E_{\rm xc}^{\rm ML-RPA}-E_{\rm xc}^{\rm RPA}\right)^2 \\
& \mathcal{L}_{2,v}^\alpha = \frac{1}{2} \frac{1}{N_e^\alpha} \frac{\Omega^\alpha}{N_{\rm spars}} \sum_{k}^{N_{\rm spars}} n(\textbf{r}_k) \left[\tilde{v}_{\rm xc}^{\rm ML-RPA}(\textbf{r}_k)-\tilde{v}_{\rm xc}^{\rm RPA}(\textbf{r}_k)\right]^2 .
\end{aligned}
\end{equation}
Here we have introduced a dimensionless weight factor $c_E$,  which allows us to balance exchange-correlation energies and potentials. Further, we normalize the loss with respect to system size via dividing by the number of electrons $N_e^{\alpha}$. Likewise, a factor $\Omega^{\alpha}/N_{\rm spars}^{\alpha}$ is included for the exchange-correlation potentials, where $\Omega^{\alpha}$ is the volume of structure $\alpha$. 
From Eq. \eqref{eq:potential shifts} it is clear that the shifted exchange-correlation potentials depend linearly on the $w_{i_B}$ just as the unshifted ones do. Thus, we can solve a system of linear equations that is obtained via minimization of the loss function \eqref{eq:L2 loss} with respect to the $w_{i_B}$,
\begin{equation}
\begin{aligned}
\partial \mathcal{L}_2 / \partial w_{i_B} \overset{!}{=} 0
\rightarrow \sum_{i_B} \phi^{\alpha}_{j,i_B} w_{i_B} = y_{j}^{\alpha} .
\end{aligned}
\end{equation}
Following \myciteauthor{Verdi2021},\cite{Verdi2021} we regularize the solution of this linear problem via pseudo inverse of the design matrix $\phi$, smoothly cutting off smaller singular values $\sigma_i$,
\begin{equation}
\sigma_i^{-1} \mapsto \frac{\sigma_i}{\sigma_i^2+(t_{\rm SVD}\sigma_{\rm max})^2} , 
\end{equation}
where we multiply the Tikhonov parameter $t_{\rm SVD}$ by the largest singular value $\sigma_{\rm max}$. Thus, $t_{\rm SVD}$ is dimensionless and we can more easily compare numerical values of $t_{\rm SVD}$ for different databases. Finally, all ML-RPA hyperparameters are summarized in Table \ref{tab:hyperparameters}.

\begin{table}[!htb] 
\caption{Table of ML-RPA hyperparameters.}\label{tab:hyperparameters}
\begin{tabular}{l @{\qquad} l @{\qquad} l} 
\hline\hline\\\\[-4.\medskipamount]
parameter & description & value \\
\hline \\\\[-4.\medskipamount]
$R_{\rm cut}$ & cutoff radius & 1.5 \AA \\
$N_{\rm rad}$ & number of radial basis functions & 4  \\
$N_{\rm des}$ & number of density descriptors & 4+4=8 \\
$\sigma$ & width of Gaussian kernel & 3.0 $e \text{\AA}^{-3}$ \\
$\sigma^{(3)}$ & weight for three-body descriptors & 1.0 \AA \\
$c_E$ & fit weight for xc-energies & 0.999 \\
$t_{\rm SVD}$ & Tikhonov regularization & $1.0 \times 10^{-9}$ \\
\hline\hline 
\end{tabular}
\end{table}

\section{Data sparsification}\label{App:Riemelmoser2023_C}

To reduce computational cost, the exchange-correlation potential is fitted not on the entire real-space grid but rather at selected representative points $\textbf{r}_k$. These points are represented  as red crosses in Fig. 1 in the main text. For each individual structure $\alpha$, we choose $N_{\rm spars}$ points via k-means sparsification. The k-means algorithm uses a metric that quantifies the similarity between density descriptors at points $\textbf{r}$ and $\textbf{r}'$. It is convenient to use the metric $d[\textbf{X}(\textbf{r}),\textbf{X}(\textbf{r}')] $ that is induced by the Gaussian kernel, $k[\textbf{X}(\textbf{r}),\textbf{X}(\textbf{r}')]$, 
\begin{equation}
\begin{aligned}
d[\textbf{X}(\textbf{r}),\textbf{X}(\textbf{r}')] 
&=k[\textbf{X}(\textbf{r}),\textbf{X}(\textbf{r})] + k[\textbf{X}(\textbf{r}'),\textbf{X}(\textbf{r}')]- 2k[\textbf{X}(\textbf{r}),\textbf{X}(\textbf{r}')] \\
&= 2 - \exp\left\{-\frac{[\textbf{X}(\textbf{r})-\textbf{X}(\textbf{r}')]^2}{2\sigma^2}\right\} .
\end{aligned}
\end{equation}
The k-means centers are initialized via farthest point sampling similar to Ref. \onlinecite{Arthur2006}. In further iterations, the centers are updated as averages over all points belonging to their respective clusters as in the standard k-means algorithm. Those points are assigned to the clusters based again on the kernel induced metric.  

Next, we combine the selected points from all structures and apply the sparsification again to choose the kernel control points $\textbf{X}^{i_B}$, compare blue squares in Fig. 1 in the main text. An interesting technical detail is that the $\textbf{r}_k$ are chosen as \textit{actual real-space points} closest to k-means centers, where $v^{\rm RPA}_{\rm xc}(\textbf{r})$ is available. For the selection of the $\textbf{X}^{i_B}$, however, we find it beneficial to use \textit{the k-means centers themselves}. That is, the chosen kernel control points correspond not to descriptors at actual real-space grid points, but rather optimized artificial ones. Integrals are evaluated on the entire real-space grid throughout.

\begin{figure}[!htb]
\centering
\includegraphics [width=0.75\linewidth,keepaspectratio=true] {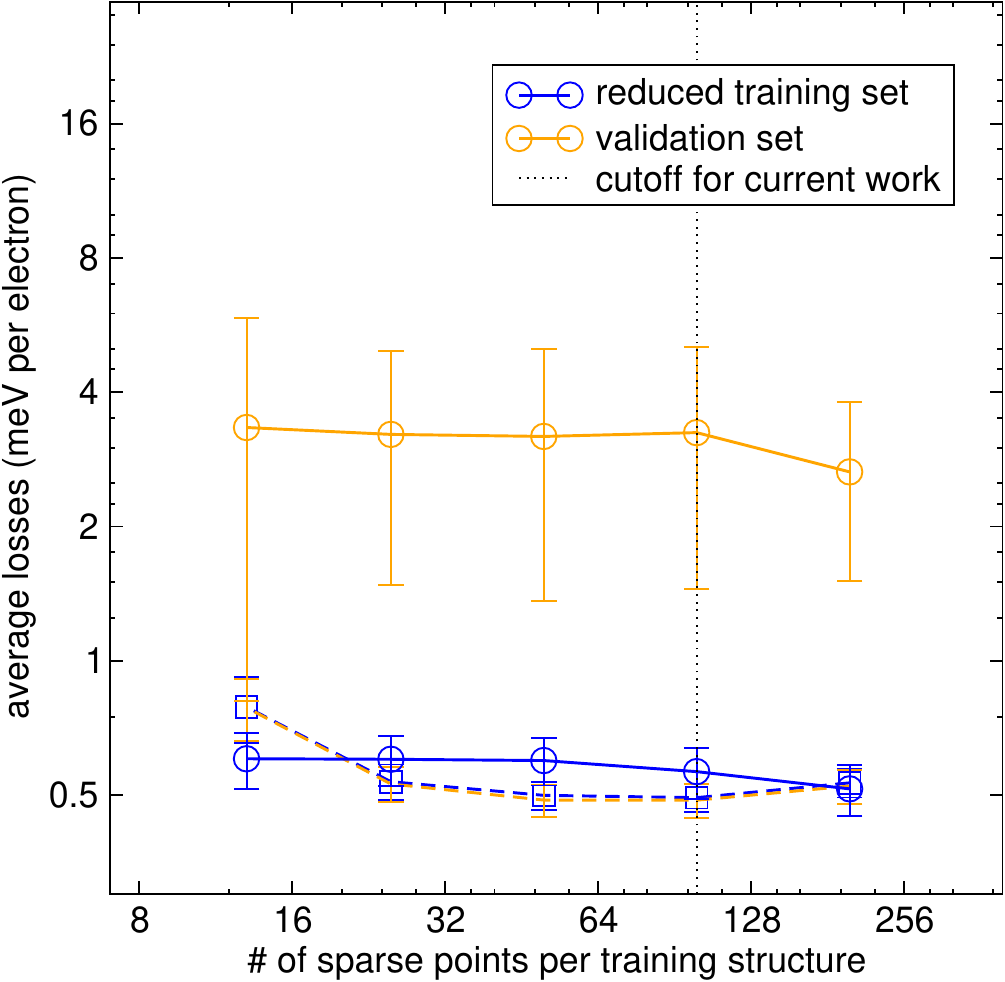}
\caption{Convergence of the ML-RPA fit with respect to $N_{\rm spars}$, that is the number of sparse points kept by the first sparsification layer (second layer fixed). Note the log-log scale. Solid lines indicate losses for exchange-correlation energies [$c_E \mathcal{L}_{E}$ in Eq. \eqref{eq:L1 losses}], and dashed lines indicate losses for exchange-correlation potentials [$(1-c_E) \mathcal{L}_{v}'$ in Eq. \eqref{eq:L1 losses}]. The losses are averaged losses over 10 random splittings (50:50).}
\label{fig:sparsification_first_layer}
\end{figure}

\begin{figure}[!htb]
\centering
\includegraphics [width=0.75\linewidth,keepaspectratio=true] {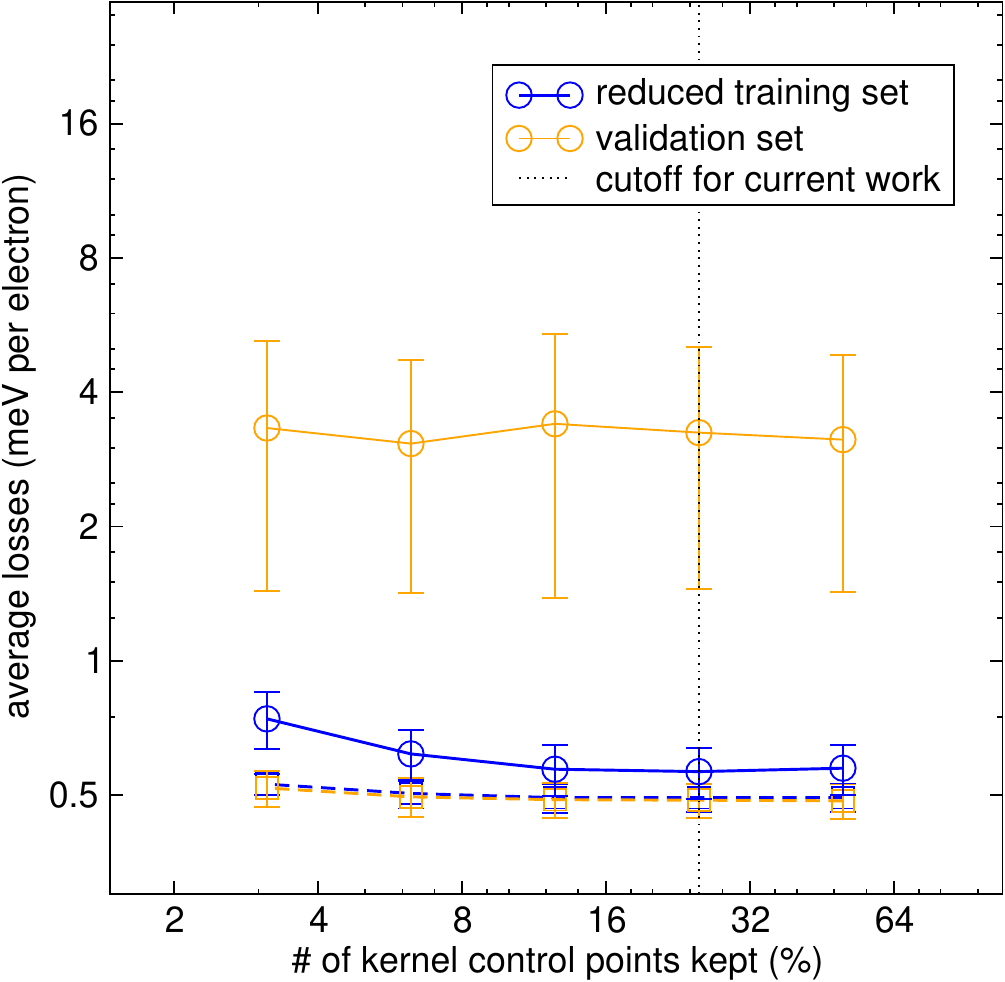}
\caption{Like Fig. \ref{fig:sparsification_first_layer}, but convergence with respect to the number of kernel control points kept by the second sparsification layer (first layer fixed).}
\label{fig:sparsification_second_layer}
\end{figure}

In the following, we demonstrate numerically the efficiency of our sparsification scheme. First, we split the ML-RPA training set randomly (50:50) into a reduced training set and a validation set. Keeping one sparsification layer fixed and varying the number of k-means clusters in the other, we monitor the loss 
\begin{equation}\label{eq:L1 losses}
\begin{aligned}
&{\mathcal{L}_1}' = \frac{1}{N_{\rm struct}}\sum_{\alpha}^{N_{\rm struct}} c_E \mathcal{L}_{1,E}^{\alpha} + (1-c_E){\mathcal{L}_{1,v}^{\alpha}}' \\
&\mathcal{L}_{1,E}^{\alpha} = \frac{1}{N_e^\alpha} |E^{\rm ML-RPA}_{\rm xc} -E_{\rm xc}^{\rm RPA} | \\
&{\mathcal{L}_{1,v}^{\alpha}}' = \frac{1}{N_e^\alpha}
\int \text{d} \textbf{r}\; n(\textbf{r}) | \tilde{v}_{\rm xc}^{\rm ML-RPA}(\textbf{r}) - \tilde{v}_{\rm xc}^{\rm RPA}(\textbf{r})|
\end{aligned}
\end{equation}
for structures in the reduced training and validation sets.
Note that the loss ${\mathcal{L}_{1,v}^{\alpha}}'$ includes the exchange-correlation potential \textit{at all real-space grid points}, thus some amount of interpolation is required to minimize ${\mathcal{L}_{1,v}^{\alpha}}'$ even for structures $\alpha$ on which ML-RPA has been trained on. This means that ${\mathcal{L}_{1,v}}'$ is less prone to overfitting and statistical error. Likewise, atomic forces in MLFF are known to be less prone to overfitting and statistical errors than energies.
Fig. \ref{fig:sparsification_first_layer} shows that in the first layer, we can downsample the number of real-space grid points per training structure ($N_{\rm spars}$) from $\mathcal{O}(10^5)-\mathcal{O}(10^6)$ to a mere 100 without loosing significant fit accuracy. The second k-means layer inputs the combined $N_{\rm struct}\times N_{\rm spars}$ points from the first layer. Fig. \ref{fig:sparsification_second_layer} shows that the number of kernel control points can be reduced by an additional factor of 4 without loss of accuracy. The other ML-RPA hyperparameters listed in Table \ref{tab:hyperparameters} were optimized in a similar fashion, minimizing validation set losses and monitoring the stability of electronic self-consistency.


\clearpage

\section{ML-RPA training set}\label{App:Riemelmoser2023_D}

\begin{figure}[!htb]
\centering
\includegraphics [width=0.75\linewidth,keepaspectratio=true] {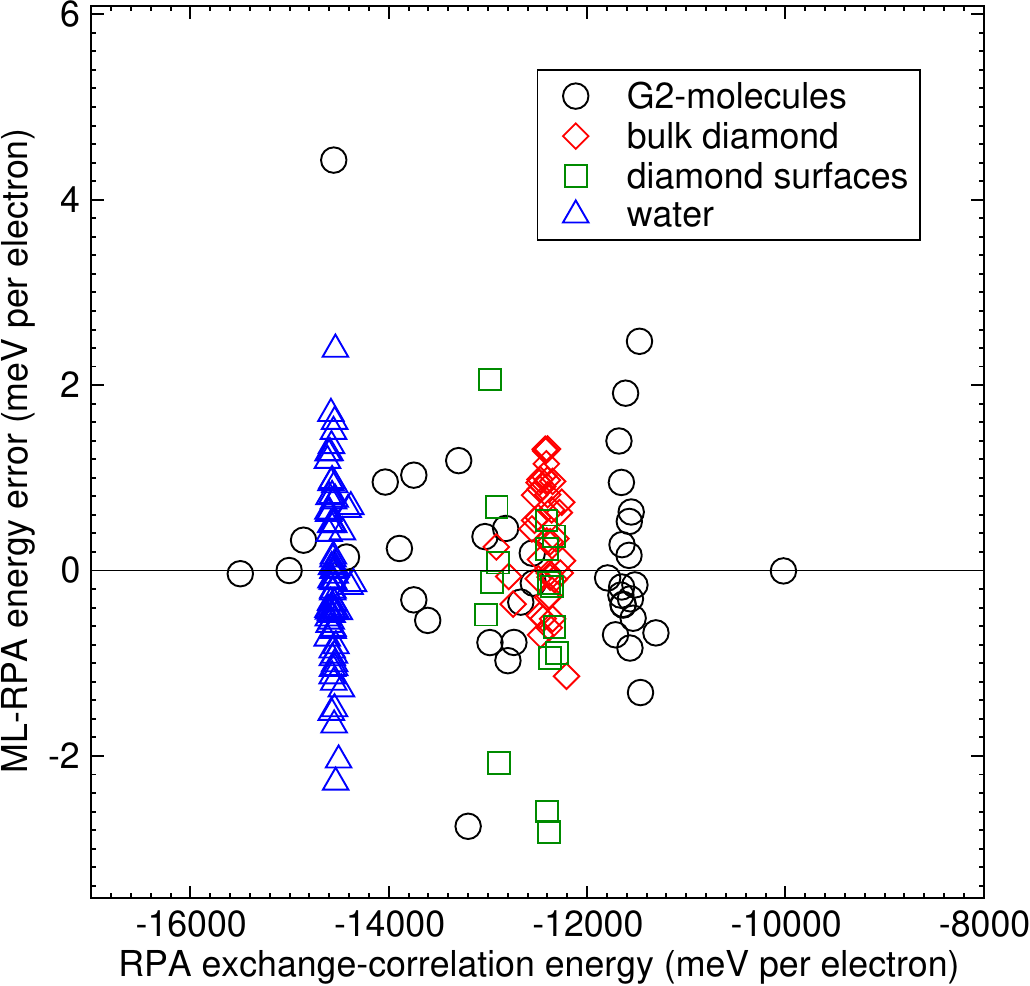}
\caption{ML-RPA energy fit error, $E_{\rm xc}^{\rm ML-RPA}/N_e-E_{\rm xc}^{\rm RPA}/N_e$. Symbols distinguish between different sub-groups of the training set. The data spread along the y-axis indicates that ML-RPA balances the fit well across the different sub-groups. Further, the spread along the x-axis shows that the G2-molecules contain very different chemical environments, whereas the other training data are more homogeneous. The $\text{H}_2\text{O}$ monomer at the experimental geometry is listed as a G2-molecule.}
\label{fig:training set energies}
\end{figure}

As a baseline, the ML-RPA training set contains 41 small molecules from the G2-2 database.\cite{Curtiss1997} We include all non-spin-polarized molecules containing elements C, O and H except for $\text{CH}_2$, which does not posses a complete octet structure. We use experimental geometries where available, and otherwise geometries from accurate quantum chemistry calculations.\cite{Johnson2002} The molecules can be further classified as 20 hydrocarbons, 19 oxygen substituted hydrocarbons, and 2 inorganic molecules ($\text{H}_2$ and $\text{O}_3$). Thus, by including the G2 molecules we sample basic bonding motives and vacuum regions. 
To train ML-RPA for our specific applications, we supplement the training set with diamond and water structures, as discussed in the main text. Table \ref{tab:training set losses} and Fig. \ref{fig:training set energies} detail the ML-RPA fit errors with respect to the different training sub-groups. 

\begin{table}[!htb] 
\caption{Average training set losses calculated via Eq. \eqref{eq:L1 losses} (in meV per electron) specified for different sub-groups of the ML-RPA training set. To compensate for the small amount of surface data, surfaces are included twice in the training set, giving them higher fit weight. The $\text{H}_2\text{O}$ monomer at the experimental geometry is listed as a G2-molecule.}\label{tab:training set losses}
\begin{tabular}{l @{\qquad} r @{\qquad} r} 
\hline\hline \\\\[-4.\medskipamount]
& $c_E \mathcal{L}_{1,E}$ & $(1-c_E){\mathcal{L}_{1,v}}'$ \\
\hline \\\\[-4.\medskipamount]
41 G2-molecules & $0.73$ & 0.30  \\
40 bulk diamond structures & $0.59$ & 0.48 \\
16 diamond surfaces (x2) & $0.93$ & 1.09 \\
76 water structures & 0.71 & 0.27 \\
\hline \\\\[-4.\medskipamount]
189 structures in total & 0.73 & 0.46 \\
\hline\hline 
\end{tabular}
\end{table}

\subsection{Diamond surfaces} 

\begin{table}[!htb] 
\caption{List of the 28 diamond surfaces used for our surface energy benchmark (see Table 2 in the main text). The first column defines the surface symmetry and stoichiometry (corresponding to the chemisorption of 1 ML hydrogen for hydrogenated surfaces, and 1 ML oxygen for oxygenated surfaces). The second column describes the geometry of the surface termination and quotes literature references, where the surfaces are characterized. Further columns show surface formation energies [see Eq. 12 in the main text] that are given in eV per surface atom. ``Exact RPA'' is the ground truth for ML-RPA, and basis set extrapolated RPA formation energies are obtained using Eq. \eqref{eq:RPA extrapolation}. Formation energies calculated with the PBE functional are also listed for comparison. All (100) and (111)-1db surfaces are included in the ML-RPA training set, all (110) and (111)-3db surfaces are out-of training.  Underlined values correspond to the most stable configurations for a given a surface termination and orientation.}\label{tab:surfaces list}
\begin{tabular}{llrrrrr}
\hline\hline \\\\[-4.\medskipamount]
& & PBE & \multicolumn{3}{c}{RPA}\\
\cline{4-6} \\\\[-4.\medskipamount]
	&  &    & extrap.  &  exact  & ML  \\
\hline \\\\[-4.\medskipamount]
\textbf{(100)*} &       &       &        &      \\  
$1\times1$        & bulk terminated, as cut\cite{Furthmueller1996} \hspace{85pt}    &   3.49              & 3.70              &   3.66               &  3.57   \\
$1\times 1$       & bulk terminated, relaxed\cite{Furthmueller1996}    &   3.36              & 3.62              &   3.59               &  3.51   \\
$1\times 1$:H    & bulk terminated, on-top\cite{Furthmueller1996}      &   0.90              & 1.21              &   1.21               &  1.16   \\
$1\times 1$:O    & bulk terminated, ketone\cite{Sque2006}              &   2.29              & 2.38              &   2.38               &  2.39   \\
$1\times 1$:O    & bulk terminated, ether\cite{Sque2006}               &   \underline{1.97}  & \underline{2.03}  &   \underline{1.99}   &  \underline{1.91}   \\
$2\times 1$       & dimer \cite{Sque2006}                              &   \underline{1.89}  & \underline{2.08}  &   \underline{2.08}   &  \underline{2.03}   \\
$2\times 1$:2H   & dimer, on-top\cite{Sque2006}                        &   \underline{0.01}  & \underline{0.20}  &   \underline{0.20}   &  \underline{0.11}   \\
$2\times 1$:2O   & dimer, bridge\cite{Zheng1992}                       &   3.45              & 3.50              &   3.48               &  3.40    \\
\hline \\\\[-4.\medskipamount]                                                                                                              
\textbf{(110)} &       &       &        &      \\                                                                                           
$1\times1$        & bulk terminated, as cut\cite{Kern1997}             &   1.90              & 2.11              &   2.09               &    1.95  \\
$1\times 1$       & bulk terminated, relaxed\cite{Kern1997}            &  \underline{1.45}   & \underline{1.69}  &   \underline{1.70}   &  \underline{1.59}  \\
$1\times 1$:H    & bulk terminated, on-top\cite{Kern1997}              &  \underline{-0.23}  & \underline{-0.12} &  \underline{-0.11}   &   \underline{-0.12}   \\
$1\times 1$:O    & bulk terminated, on-top$^\dagger$                   &   3.38              & 3.61              &   3.60               &    3.58   \\
\hline \\\\[-4.\medskipamount]                                                                                                
\multicolumn{6}{l}{$^*$included in the ML-RPA training set} \\
\hline \\\\[-4.\medskipamount]
\multicolumn{6}{l}{$^\dagger$similar to the hydrogenated (110) surface, the calculated C-O bond length is $1.36 \text{ \AA}$.} \\
\hline\hline
\end{tabular}
\end{table}

\begin{table*}[!htb] 
\caption*{Table \ref{tab:surfaces list}, continued.}
\begin{tabular}{llrrrrr}
\hline\hline \\\\[-4.\medskipamount]
& & PBE & \multicolumn{3}{c}{RPA}\\
\cline{4-6} \\\\[-4.\medskipamount]
	&  &    & extrap.  &  exact  & ML  \\
\hline \\\\[-4.\medskipamount]                                                                                                                                        
\textbf{(111)-1db*} &       &       &        &      \\                                                                                      
$1\times1$        & bulk terminated, as cut\cite{Kern1996}             &   2.46              &  2.71             &   2.69               &   2.52    \\
$1\times 1$       & bulk terminated, relaxed\cite{Kern1996}            &   1.94              &  2.30             &   2.29               &   2.12    \\
$1\times 1$:H    & bulk terminated, on-top\cite{Kern1996}              &  \underline{-0.34}  & \underline{-0.19} &  \underline{-0.19}   &  \underline{-0.24}    \\
$1\times1$:O     & bulk terminated, on-top\cite{Loh2002}               &  \underline{ 2.84}  &  3.19             &   3.19               &   2.99    \\
$2\times 1$:2O  & bulk terminated (distorted), peroxide\cite{Loh2002}  &   2.99              &  \underline{3.02} &   \underline{2.99}   &   \underline{2.94}   \\
$2\times1$        & Pandey chain\cite{Kern1996}                        &  \underline{1.18}   &  \underline{1.43} &   \underline{1.41}   &   \underline{1.34}   \\
$2\times1$:2H   & Pandey chain, on-top\cite{Kern1996}                  &   0.31              &  0.49             &   0.47               &   0.40   \\
$2\times 1$:2O  & Pandey chain, ketone\cite{Loh2002}                   &   2.94              &  3.08             &   3.09               &   2.99   \\
\hline \\\\[-4.\medskipamount]                                                                                                              
\textbf{(111)-3db} &       &       &        &      \\                                                                                       
$1\times1$        & bulk terminated, as cut\cite{Kern1996a}            &   4.27              & 4.41              &   4.37               &   4.37  \\
$1\times 1$       & bulk terminated, relaxed\cite{Kern1996a}           &   4.26              & 4.41              &   4.37               &   4.37  \\
$1\times 1$:H    & bulk terminated, on-top\cite{Kern1996a}             &   3.01              & 3.38              &   3.36               &   3.43  \\
$1\times1$:O   & bulk terminated, ketone\cite{Zheng1992}               &   3.50              & 3.92              &   3.90               &   3.87   \\
$2\times1$        & Seiwatz chain\cite{Kern1996a}                &   \underline{2.44}  & \underline{2.71}  &   \underline{2.70}   &   \underline{2.56}   \\
$2\times1$:2H   & Seiwatz chain, on-top\cite{Kern1996a}                &   \underline{0.00}  & \underline{0.17}  &   \underline{0.17}   &   \underline{0.11}   \\
$2\times1$:2O   & Seiwatz chain, ketone\cite{Zheng1992}                &   \underline{2.67}  & \underline{2.91}  &   \underline{2.90}   &   \underline{2.79}   \\
\hline \\\\[-4.\medskipamount]                                                                                                
\multicolumn{6}{l}{$^*$included in the ML-RPA training set} \\
\hline\hline
\end{tabular}
\end{table*}

Table \ref{tab:surfaces list} specifies the 28 diamond surfaces used to benchmark different DFT functionals (see Table 2 in the main text). These surfaces have been described in detail in past studies, we refer to the original references for more complete descriptions of the surface geometries. \cite{Kern1996a,Kern1997,Loh2002,Chaudhuri2022,Zheng1992} In the following, we briefly comment on the interesting case of oxygenated (111) surfaces, where several (meta-)stable configurations exist that are close in energy. The least stable surface is the (111)-3db symmetric ($1\times1$) oxygenated surface, while the most stable surface is the $2\times1$ reconstructed (111)-3db oxygenated chain surface, and the three oxygenated (111)-1db surfaces are in between.  The (111)-3db oxygenated chain surface can be interpreted as clean (111)-1db surface adsorbing a monolayer of CO molecules, compare Fig. 4 in the main text. This is significant insofar as CO molecules have been reported to be the main desorption product in temperate-programmed desorption experiments on oxygenated (111) surfaces.\cite{Loh2002} The C-O bond length of the oxygenated (111)-3db chain surface is calculated to be $1.20 \text{ \AA}$ (we use PBE geometries throughout). This clearly indicates strong double bonds, in comparison, the C-O bond lengths for the (111)-1db oxygenated surfaces are $1.20 \text{ \AA}$ (ketone) and $1.31 \text{ \AA}$ (on-top), as well as $\{1.43 \text{ \AA}, 1.44 \text{ \AA}\}$ (peroxide). 
Out of the three oxygenated (111)-1db surfaces, exact RPA and ML-RPA predict the peroxide structure to be the most stable one, whereas PBE favors the on-top configuration. Otherwise, PBE, ML-RPA and exact RPA predict identical energy orderings throughout, see underlined values in Table \ref{tab:surfaces list}. 

Finally, correlated methods such as RPA are known to converge slowly with respect to the basis set size.\cite{Furche2001,Klimes2014a} To confirm the technical converge of our RPA calculations with respect to basis set, we extrapolate the RPA correlation energies with respect to the energy cutoff $E^{\chi}_{\rm max}$ using the formula\cite{Harl2010,Riemelmoser2020}

\begin{equation}\label{eq:RPA extrapolation}
E_{\rm c}^{\rm RPA} (E^{\chi}_{\rm max}) = E_{\rm c}^{\rm RPA} (E^{\chi}_{\rm max}=\infty) + \frac{\rm const}{{E^{\chi}_{\rm max}}^{3/2}} .
\end{equation}

Table \ref{tab:surfaces list} shows that the extrapolated formation energies are slightly larger than our ``exact RPA'' values, which form the ground truth for ML-RPA. The agreement between ``exact'' and extrapolated RPA formation energies is 40 meV or better throughout.  

The basis set incompleteness error also causes a slight underbinding of bulk diamond. The extrapolated values for the equlibrium lattice constant is $3.572$ \AA, whereas ``exact RPA'' predicts $3.581$ \AA.

\section{Machine learning force fields for liquid water}\label{App:Riemelmoser2023_E}

\noindent RPA-MLFF is trained directly on total energies and forces from RPA calculations. Since RPA force calculations are very expensive, we use a cheaper RPA setup here. Specifically, an energy cutoff of 500 eV is used for the plane-wave basis sets expanding both the orbitals and the response function (ML-RPA ground truth calculations use 600 eV and 400 eV, respectively). Our tests, however, indicate that this difference has only minor effects on the predicted liquid water RDFs. 
The RPA-MLFF training data set consists 107 water structures containing 32 molecules or less. We use a compact hyperparameter setup that has shown to enable an efficient and accurate MLFF training for liquid water.\cite{Jinnouchi2023} In particular, the radial and angular descriptors are separated as described in Ref. \onlinecite{Jinnouchi2020}, and the angular descriptors are truncated at angular momentum number $l=2$. Further, the radial descriptors use a cutoff of $6.0 \text{ \AA}$ and 8 radial basis functions, whereas the angular descriptors use a smaller cutoff of $4.0 \text{ \AA}$ and only 6 radial basis functions. For a general description of the MLFF scheme see also Refs. \onlinecite{Jinnouchi2019} and \onlinecite{Jinnouchi2019a}. The RPA-MLFF training set errors are given in Table \ref{tab:MLFF losses}.

The MLFF used to speed up ML-RPA uses the same descriptors as RPA-MLFF, but is trained on-the-fly\cite{Jinnouchi2019,Jinnouchi2019a} at fixed volume using a temperature ramp from 270 K to 370 K and a supercell containing 64 water molecules. In this way, a training set of around 400 structures is created. Thus, the combination of ML-RPA and MLFF allows for significantly more MLFF ground truth calculations, since ML-RPA is orders of magnitude cheaper than exact RPA. Moreover, the stresses predicted by ML-RPA can be seamlessly included in the MLFF training, whereas this would not be as easily possible for RPA-MLFF (RPA stress tensors can be included via finite differences,\cite{Liu2022} but this is very expensive). The MLFFs for PBE+TS and RPBE+D3(BJ) are also trained on-the-fly, fit accuracy being similar to ML-RPA (see Table \ref{tab:MLFF losses}). 

The production run which produced the RDFs shown in Fig. 5 in the main text used 200000 MD steps with a time step of $\text{1.5 fs}$. To increase sampling efficiency, the mass of the hydrogen atom was increased by a factor 8 (this does not affect static properties such as the RDF). For PBE+TS, which diffuses very weakly at 300 K, we further increased the supercell size to 512 water molecules, and increased the length of the MD run by a factor 3. Next, we define an independent test set of 10 liquid water snapshots sampled from the RPA-MLFF production run (64 water molecules). A low test set energy root means square error (RMSE) of 1.6 meV per atom demonstrates (i) that the ambient conditions used are well covered by the RPA-MLFF training set, (ii) that RPA-MLFF is able to extrapolate to the slightly larger simulation cell (64 water molecules). We also use this test set to evaluate the on-the-fly MLFFs, and the test set energy RMSEs are excellent (0.4 meV per atom for ML-RPA, 0.6 meV per atom for PBE+TS, and 0.5 meV per atom for RPBE+D3(BJ), respectively).

\begin{table}[!htb] 
\caption{Training set errors (root means square error, RMSE) for the MLFFs trained for liquid water. Energy RMSEs are given in meV per atom, force RMSEs are given in $\text{meV \AA}^{-1}$, and stress RMSEs are given in kbar.}\label{tab:MLFF losses}
\begin{tabular}{l @{\qquad} c @{\qquad} c @{\qquad} c @{\qquad} c} 
\hline\hline \\\\[-4.\medskipamount]
& structures & energy & force & stress  \\
\hline \\\\[-4.\medskipamount]
RPA-MLFF & 107 & 0.6 & 27.0 & ---\\  
ML-RPA-MLFF & 389 & 0.4 & 29.9 & 0.26 \\
PBE+TS-MLFF & 490 & 0.3  & 31.7 & 0.21 \\
RPBE+D3(BJ)-MLFF & 521 & 0.3 & 30.5 & 0.21  \\
\hline\hline 
\end{tabular}
\end{table}

\section{Water hexamer benchmark}\label{App:Riemelmoser2023_F}

\begin{figure}[!htb] \label{hexamers_supp}
\centering
\includegraphics [width=0.75\linewidth,keepaspectratio=true] {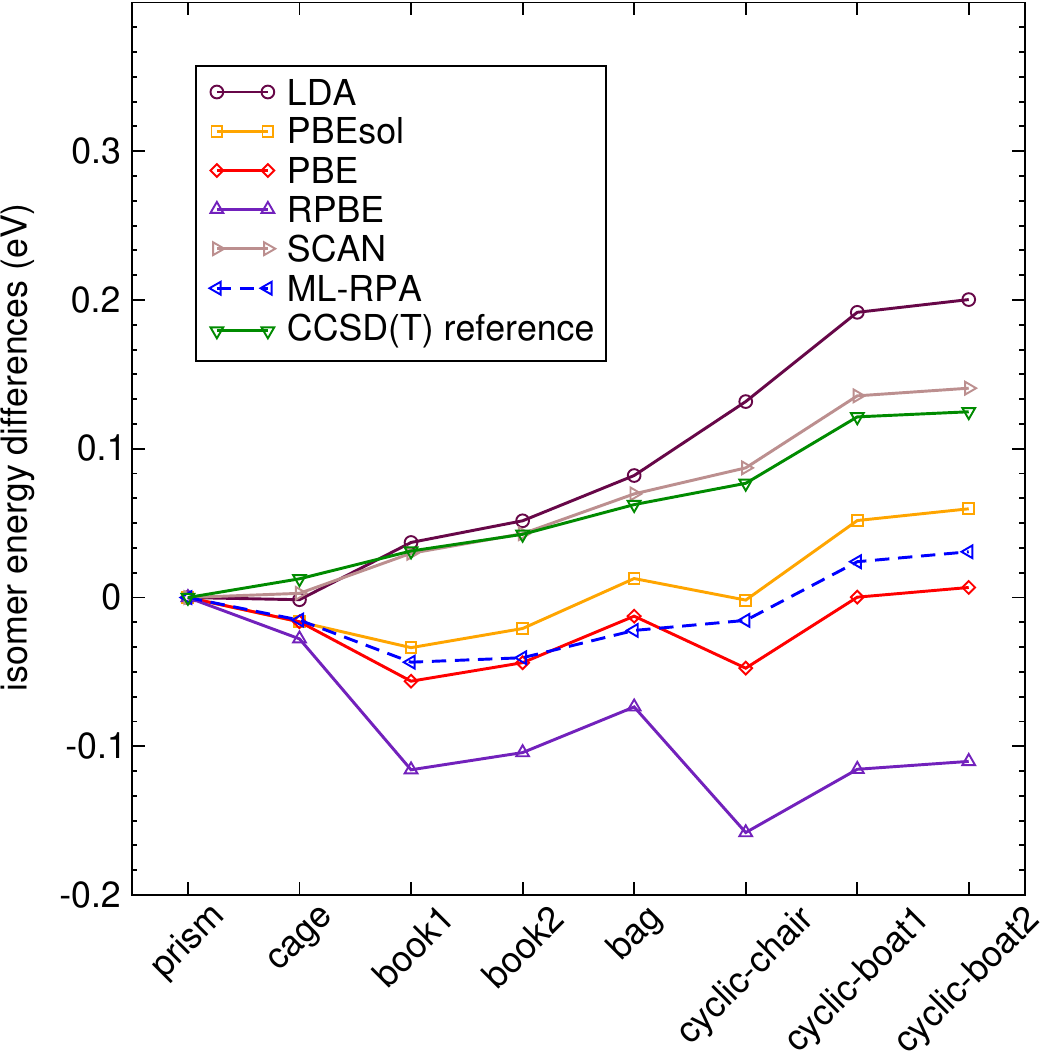}
\caption{Binding energies differences of eight hexamers as in Fig. 6 in the main text. ML-RPA results are compared to various semi-local density functionals. Lines drawn are only guides to the eye.}
\label{fig:hexamers_supp_semilocal}
\end{figure}

\begin{figure}[!htb] \label{hexamers_supp_vdW}
\centering
\includegraphics [width=0.75\linewidth,keepaspectratio=true] {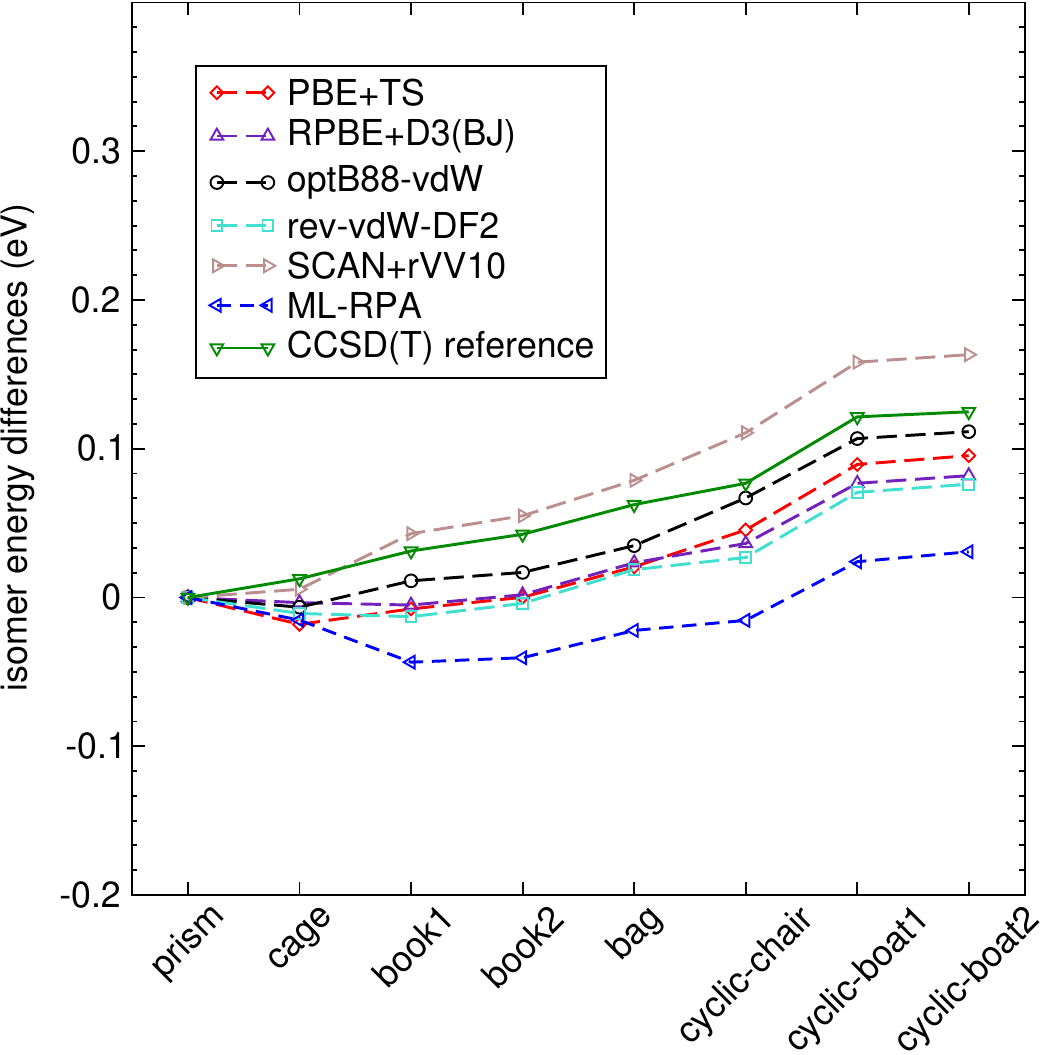}
\caption{Like Fig. \ref{fig:hexamers_supp_semilocal}, but for various vdW functionals.}
\label{fig:hexamers_supp_vdW}
\end{figure}

Figs. \ref{fig:hexamers_supp_semilocal} and \ref{fig:hexamers_supp_vdW} detail the water hexamer benchmark for various semi-local and vdW functionals, respectively. The vdW corrected PBE+TS and RPBE+D3 functionals well improve over their respective GGA base functionals, whereas SCAN+rVV10 performs slightly worse than the pristine SCAN functional.


\clearpage

\bibliography{Riemelmoser2023}